\documentclass[%
 reprint,
 amsmath,amssymb,
 aps,
floatfix,
]{revtex4-2}

\usepackage{graphicx}
\usepackage{dcolumn}
\usepackage{bm}

\begin{document}

\preprint{APS/123-QED}

\title{Exact analytical solution for density matrix of a non-equilibrium polariton Bose-Einstein condensate}

\author{Vladislav Yu. Shishkov}
\email{vladislavmipt@gmail.com}
\author{Evgeny S. Andrianov}%
\affiliation{%
 Dukhov Research Institute of Automatics (VNIIA), 22 Sushchevskaya, Moscow 127055, Russia
}%
\affiliation{%
 Moscow Institute of Physics and Technology, 9 Institutskiy pereulok, Dolgoprudny 141700, Moscow region, Russia
}%
\affiliation{
 Skolkovo Institute of Science and Technology, Moscow, Russian Federation
}

\author{Anton V. Zasedatelev}
\affiliation{
 Skolkovo Institute of Science and Technology, Moscow, Russian Federation
}%

\author{Pavlos G. Lagoudakis}
\affiliation{
 Skolkovo Institute of Science and Technology, Moscow, Russian Federation
}%
\affiliation{
 Department of Physics and  Astronomy, University of Southampton, Southampton, SO17 1BJ, United Kingdom
}%

\author{Yurii E. Lozovik}
\affiliation{%
 Institute for Spectroscopy RAS, 5 Fizicheskaya, Troitsk 142190, Russia
}%
\affiliation{%
 Moscow Institute of Electronics and Mathematics, National Research
University Higher School of Economics, 101000, Moscow, Russia
}%
\affiliation{
 Skolkovo Institute of Science and Technology, Moscow, Russian Federation
}

\date{\today}

\begin{abstract}
In this letter, we give an analytical quantum description of a non-equilibrium polariton Bose-Einstein condensate (BEC) based on the solution of the master equation for the full polariton density matrix in the limit of fast thermalization.
We find the density matrix of a non-equilibrium BEC, that takes into account quantum correlations between all polariton states. 
We show that the formation of BEC is accompanied by the build-up of cross-correlations between the ground state and the excited states reaching their highest values at the condensation threshold. 
Despite the non-equilibrium nature of polariton systems, we show the average population of polariton states exhibits the Bose-Einstein distribution with an almost zero effective chemical potential above the condensation threshold similar to an equilibrium BEC. 
We demonstrate that above threshold the effective temperature of polariton condensate drops below the reservoir temperature.

\end{abstract}

\maketitle


\textit{Introduction.} The theoretical prediction of BEC was made at the dawn of modern statistical physics~\cite{bose1924plancks, einstein1925quantentheorie}. 
However, the first experimental realization of the BEC required more than 70 years of physicists' efforts~\cite{anderson1995observation, davis1995bose}. 
BEC was first observed for atoms in a magneto-optical trap at cryogenic temperatures~\cite{anderson1995observation, davis1995bose}. 
A rapid development of semiconductor technologies made it possible to experimentally realize polariton BEC in solid state at elevated temperature~\cite{deng2003polariton, kasprzak2006bose, combescot2015excitons, byrnes2014exciton, wertz2010spontaneous, balili2007bose, estrecho2018single, klaas2018photon, sun2017bose, deng2010exciton, klaas2018photon, imamog1996nonequilibrium, keeling2020bose, wei2019low, zasedatelev2019room, plumhof2014room}. 
Polaritons are hybrid quasiparticles formed by the strong light-matter interaction of photons in a cavity with excitons of a semiconductor material or organic dyes~\cite{combescot2015excitons, kavokin2003thin}. 
Due to their small effective mass and low density of states~\cite{byrnes2014exciton, wertz2010spontaneous, deng2010exciton, imamog1996nonequilibrium}, the critical temperature for a polariton BEC can be as high as tens or even hundreds of kelvin~\cite{deng2003polariton, kasprzak2006bose, plumhof2014room}.
Such light-matter condensed states attract fundamental as well as practical interest due to their potential in all-optical manipulation and low-energy optoelectronic applications~\cite{sanvitto2016road, keeling2020bose}.

Being open dissipative systems polariton condensates intrinsically are far from equilibrium. 
To maintain the number of polaritons at a certain level an external pumping is required.
Therefore, the methods of equilibrium thermodynamics must be applied with caution to the description of non-equilibrium polariton condensates~\cite{combescot2015excitons, malpuech2002room, kavokin2017microcavities, porras2002polariton}.
The description of the dynamics of polariton condensates require more sophisticated approaches (for a review, see the Supplement in~\cite{keeling2020bose}). 
Generally, complete information about the polariton condensate can be accessed through its full density matrix. 
When dissipation processes are described in the Markov approximation, their dynamics obeys the corresponding Lindblad equation~\cite{kavokin2017microcavities}.
However, due to the large number of excited states, the master equation for the polariton density matrix cannot be solved exactly.
Previous studies largely rely on the semiclassical Maxwell--Boltzmann equations to the average population of polaritons in condensates~\cite{malpuech2002room, banyai2002real, cao2004condensation, doan2008coherence, tassone1997bottleneck} or other mean-field theories~\cite{kirton2013nonequilibrium, kirton2015thermalization, strashko2018organic}.
This approach takes into account both dissipation and pumping as well as the processes leading to the thermalization of the polaritons~\cite{kavokin2017microcavities, hartwell2010numerical}.
The Maxwell--Boltzmann equations reproduce the observed Gibbs distribution of the polaritons in the excited states~\cite{hartwell2010numerical, kasprzak2008formation}.
However, this approach does not provide any information neither on the coherence of polaritons nor on the correlations between the polariton states~\cite{kavokin2017microcavities, sanvitto2012exciton}.
Recently, a theory beyond the mean-field approximation was developed~\cite{arnardottir2020multimode}.
This theory consistently describes lasing in weakly and strongly coupled organic systems with strong exciton-vibron coupling. It takes two-time correlations into account allowing to study evolution of coherence and thermalization in such systems, albeit limited to the first-order correlations.

A way to get information about the coherence of the polariton BEC is to reduce the complete master equation to an effective one for the density matrix of the ground polariton state only, where the contribution of the excited polariton states is taken into account as a whole effectively~\cite{doan2008coherence, sanvitto2012exciton, kirton2015thermalization, rubo2003dynamical, laussy2004spontaneous, PhysRevLett.103.096404, https://doi.org/10.1002/pssa.200304065, https://doi.org/10.1002/pssc.200303205}.
Such reduced equation for the density matrix allows one to obtain complete information about the quantum state of the ground polariton state upon the BEC formation~\cite{kavokin2017microcavities, sanvitto2012exciton}. 
However, this approach does not provide any information about the coherence of the polaritons in the excited states, their statistical properties and their cross-correlations between the ground state. 

In this letter, we propose a complete quantum-mechanical approach for obtaining the density matrix of non-equilibrium polariton condensates in the limit of fast polariton thermalization. 
We find the stationary density matrix of the polariton BEC, which allows us to analyze the correlations between all polariton states. 
We show that cross-correlations between polaritons in the ground state and those in the excited states arises at the condensation threshold. 
This property is a unique feature of non-equilibrium BEC.
We show that the number of polaritons in the ground and excited states obeys Bose--Einstein distribution with effective temperature well below the reservoir temperature and an almost zero effective chemical potential above condensation threshold. 

\textit{Master equation for the lower polaritons.}
Strong light-matter interaction of the cavity photons with an optical transition of active material gives rise for new eigenstates, namely lower and upper polariton branches.
The polaritons of the upper branch and excitons~\cite{zasedatelev2019room, byrnes2014exciton}) are able to scatter towards lower polariton branch forming BEC at high enough excitation density.
There are several mechanisms allowing such energy relaxation in polariton systems. 
Scattering with electrons~\cite{lagoudakis2003electron} acoustic~\cite{kavokin2017microcavities} and high-energy optical phonons~\cite{maragkou2010longitudinal} or vibrons in organic materials \cite{zasedatelev2019room,coles2011vibrationally} are among them. 
Pair-particle scattering processes can also contribute to energy relaxation towards the low-lying polariton states at high polariton densities~\cite{savvidis2000angle}.

At the same time polaritons undergo multiple scattering at the lower polariton branch followed by thermalization and condensation above critical polariton density.
Since the non-equilibrium BEC occurs in the lower polariton branch, below we consider the dynamics of the lower polariton branch only.
We describe the energy transfer from the upper polaritons or excitons to the lower polaritons as an effective incoherent pumping which is reasonable assumption for majority of experimental realizations~\cite{kasprzak2006bose,plumhof2014room,sun2017bose}

We suppose that the polaritons with frequencies ${\omega _j}$ of the lower polariton branch near the minimum are described by bosonic creation $\hat a_j^\dag $ and annihilation $\hat a_j$  operators~\cite{kavokin2017microcavities, sanvitto2012exciton} which obey the commutation relation $\left[ {{{\hat a}_j},\hat a_{j'}^\dag } \right] = {\delta _{jj'}}$. 
In this case, the Hamiltonian of the polaritons takes the form 
\begin{equation}\label{Hamiltonian}
{\hat H_{{\rm{LP}}}} = \sum\limits_{j = 0}^M {{\omega _j}\hat a_j^\dag {{\hat a}_j}} 
\end{equation}
The state with $j=0$ is the ground state.

We describe the dynamics of the polaritons in the lower polariton branch through the density matrix $\hat \rho$. 
We consider the relaxation and pumping processes in the Born--Markov approximation~\cite{breuer2002theory}. 
In such a case, the density matrix is governed by the master equation in the Lindblad form. 
The dissipation of the lower polaritons is described by the Lindblad superoperator
\begin{equation}\label{Lindblad_diss}
{L_{{\rm{diss}}}}\left( {\hat \rho } \right) = \sum\limits_{j = 0}^M {{\gamma _j}\left( {{{\hat a}_j}\hat \rho \hat a_j^\dag  - \frac{1}{2}\hat \rho \hat a_j^\dag {{\hat a}_j} - \frac{1}{2}\hat a_j^\dag {{\hat a}_j}\hat \rho } \right)} 
\end{equation}
where ${\gamma _j}$ is the dissipation rate of the $j$-th state. 
The effective incoherent pumping of the lower polaritons can be described by the Lindblad superoperator
\begin{multline}\label{Lindblad_pump}
{L_{{\rm{pump}}}}\left( {\hat \rho } \right) = \sum\limits_{j = 0}^M {{\kappa _j}\left( {{{\hat a}_j}\hat \rho \hat a_j^\dag  - \frac{1}{2}\hat \rho \hat a_j^\dag {{\hat a}_j} - \frac{1}{2}\hat a_j^\dag {{\hat a}_j}\hat \rho } \right)} \\ + \sum\limits_{j = 0}^M {{\kappa _j}\left( {\hat a_j^\dag \hat \rho {{\hat a}_j} - \frac{1}{2}\hat \rho {{\hat a}_j}\hat a_j^\dag  - \frac{1}{2}{{\hat a}_j}\hat a_j^\dag \hat \rho } \right)} 
\end{multline}
where ${\kappa _j}$ is the pumping rate of the $j$-th state. 
The operator ${L_{{\rm{pump}}}}\left( {\hat \rho } \right)$ leads to the following dynamics of the average polariton population in the $j$-th state $\left\langle {\hat a_j^\dag {{\hat a}_j}} \right\rangle$ and the average polariton amplitude in the $j$-th state $\left\langle {{{\hat a}_j}} \right\rangle$: ${\left( {{{d\left\langle {\hat a_j^\dag {{\hat a}_j}} \right\rangle } \mathord{\left/
 {\vphantom {{d\left\langle {\hat a_j^\dag {{\hat a}_j}} \right\rangle } {dt}}} \right.
 \kern-\nulldelimiterspace} {dt}}} \right)_{{\rm{pump}}}} = {\kappa _j}$ and ${\left( {{{d\left\langle {{{\hat a}_j}} \right\rangle } \mathord{\left/
 {\vphantom {{d\left\langle {{{\hat a}_j}} \right\rangle } {dt}}} \right.
 \kern-\nulldelimiterspace} {dt}}} \right)_{{\rm{pump}}}} = 0$.
Therefore, the action of incoherent pumping~(\ref{Lindblad_pump}) leads to the excitation of a certain number of polaritons per unit time in the corresponding state, without affecting their phase~\cite{scully1999quantum}. 

Thermalization of the lower polaritons may occur due to different physical processes depending on the system.
For example, in organic polariton systems thermalization occurs due to their nonlinear interaction with low frequency vibrations~\cite{strashko2018organic, litinskaya2004fast, mazza2009organic, bittner2012estimating, coles2011vibrationally, cwik2014polariton, somaschi2011ultrafast, ramezani2018nonlinear}. 
For polariton states in inorganic semiconductors thermalization predominantly goes through interactions with acoustic phonons or free charges~\cite{kavokin2017microcavities}.
Also the thermalization of the polaritons can occur due to polariton-polariton scattering~\cite{deng2010exciton}.
However, regardless of the mechanism, this thermalization can be described through the Lindblad superoperator~\cite{kavokin2017microcavities, sanvitto2012exciton}.
\begin{multline}\label{Lindblad_term}
{L_{{\rm{therm}}}}\left( {\hat \rho } \right) = \\
\sum\limits_{j = 0}^M {\sum\limits_{k = 0}^M {{\Gamma _{jk}}\left( {{{\hat a}_j}\hat a_k^\dag \hat \rho {{\hat a}_k}\hat a_j^\dag  - \frac{1}{2}\hat \rho {{\hat a}_k}\hat a_j^\dag {{\hat a}_j}\hat a_k^\dag  - \frac{1}{2}{{\hat a}_k}\hat a_j^\dag {{\hat a}_j}\hat a_k^\dag \hat \rho } \right)} }
\end{multline}
where ${\Gamma _{jk}}$ is the transition rate from the $j$-th polariton state to the $k$-th state. 
The thermalization rates ${\Gamma _{jk}}$ obey the Kubo--Martin--Schwinger relation ${\Gamma _{jk}}/{\Gamma _{kj}} = \exp \left( {\left( {{\omega _j} - {\omega _k}} \right)/T} \right)$, where $T$ is the temperature of intermolecular oscillations of the organic dyes or the temperature of the phonons in the semiconductors. 

Thus, the complete master equation for the density matrix of the polaritons at the lower polariton branch $\hat \rho$ has the form
\begin{equation}\label{Master_equation}
{\frac{\partial \hat \rho }{\partial t}} = \frac{i}{\hbar }\left[ {\hat \rho ,{{\hat H}_{{\rm{LP}}}}} \right] + {L_{{\rm{diss}}}}\left( {\hat \rho } \right) + {L_{{\rm{pump}}}}\left( {\hat \rho } \right) + {L_{{\rm{therm}}}}\left( {\hat \rho } \right)
\end{equation}
Usually it is difficult to solve the master equation~(\ref{Master_equation}). 
For instance, if we take into account $M + 1$ polariton states and $N$ excitations in each state, then the total number of differential equations given by~(\ref{Master_equation}) is ${\left( {N + 1} \right)^{2\left( {M + 1} \right)}}$.
Here we consider the fast thermalization limit and show that in this case the master equation~(\ref{Master_equation}) can be reduced to $\left( {N + 1} \right)$ differential equations.

Below we describe our approach in two stages.
First, we find all possible thermalized density matrices of the polaritons. 
We assume that thermalization is the fastest process in the system, i.e., ${\Gamma _{0j}} (1 + \langle \hat a^\dag_0\hat a_0\rangle) \gg {\kappa _j},{\gamma _j}$ (see Discussion). In such a case, at times $t \ll \gamma _j^{ - 1},\kappa _j^{ - 1}$ the density matrix obeys the approximate differential equation 
\begin{equation}\label{Diff_Fast_termalization}
\frac{{d\hat \rho }}{{dt}} = \frac{i}{\hbar }\left[ {\hat \rho ,{{\hat H}_{{\rm{LP}}}}} \right] +{L_{{\rm{therm}}}}\left( {\hat \rho } \right)
\end{equation}
After the time ${\Gamma _{0j}}^{-1}(1+\langle \hat a^\dag_0\hat a_0\rangle)^{ - 1} \ll t \ll \gamma _j^{ - 1},\kappa _j^{ - 1}$, the system reach its quasistationary state which is determined by the following equation:
\begin{equation}\label{Fast_termalization}
\frac{i}{\hbar }\left[ {\hat \rho ,{{\hat H}_{{\rm{LP}}}}} \right] +{L_{{\rm{therm}}}}\left( {\hat \rho } \right) = 0
\end{equation}
In the limits ${\kappa _j}/{\Gamma _{0j}}(1+\langle \hat a^\dag_0\hat a_0\rangle) \to 0$ and ${\gamma _j}/{\Gamma _{0j}}(1+\langle \hat a^\dag_0\hat a_0\rangle) \to 0$, we can suppose that the system reaches quasistationary state at instantaneous time that is defined by the condition~(\ref{Fast_termalization}).
Applying the theory~\cite{shishkov2018zeroth}, we can obtain the general form of the density matrix of this quasistationary state.
Indeed, the relaxation operator ${L_{{\rm{therm}}}}\left( {\hat \rho } \right)$, defined by the expression~(\ref{Lindblad_term}), conserves the total number of lower polaritons~\cite{kavokin2017microcavities}. 
Therefore the operator of total number of lower polaritons $\sum_{j = 0}^M\hat a^\dag_j\hat a_j$ is the integral of motion for the thermalization process.
According to~\cite{shishkov2018zeroth}, the presence of this integral of motion implies that the system has invariant subspaces $\left| {{n_0},{n_1},...,{n_M}} \right\rangle \left\langle {{n_0},{n_1},...,{n_M}} \right|$ with the total number of polaritons equal to $\sum\nolimits_{j = 0}^M {{n_j}} = N$.
Being in the invariant subspace at initial moment in time, the system stays in this invariant subspace in the subsequent moments.
In each invariant subspace the Gibbs distribution is established with a temperature $T$~\cite{shishkov2018zeroth}.
The corresponding density matrix is $Z_N^{ - 1}\sum\nolimits_{{n_0} + ... + {n_M} = N} {w_0^{{n_0}}...w_M^{{n_M}}\left| {{n_0},...,{n_M}} \right\rangle \left\langle {{n_0},...,{n_M}} \right|}$, where ${w_j} = \exp \left( {\left( {{\omega _0} - {\omega _j}} \right)/T} \right)$, and ${Z_N}$ is the partition function given that the total number of polaritons is $N$
\begin{equation}\label{Stat_sum}
{Z_N} = \sum\limits_{{n_0} + ... + {n_M} = N} {w_0^{{n_0}}...w_M^{{n_M}}} 
\end{equation}
Some properties of the partition function ${Z_N}$ are discussed in the Supplemental Material~\ref{sec:SI_1}. 
The general expression for the density matrix $\hat \rho \left( t \right)$ that fulfills the condition~(\ref{Fast_termalization}) is the sum of the Gibbs distributions in each invariant subspace taken with the coefficients ${P_N}\left( t \right)$
\begin{multline}\label{Density_matrix}
\hat \rho \left( t \right) = \sum\limits_{N = 0}^{ + \infty } {P_N}\left( t \right)\frac{1}{{{Z_N}}}\sum\limits_{{n_0} + ... + {n_M} = N} {w_0^{{n_0}}...w_M^{{n_M}}} \\
\times{\left| {{n_0},...,{n_M}} \right\rangle \left\langle {{n_0},...,{n_M}} \right|}      
\end{multline}
The coefficients ${P_N}\left( t \right)$ are the probabilities that there are $N$ polaritons in total in the low polariton branch, consequently $\sum\nolimits_{N = 0}^{ + \infty } {{P_N}\left( t \right)}  = 1$.

On the second stage, we substitute the general thermalized density matrix~(\ref{Density_matrix}) into the master equation~(\ref{Master_equation}) to account for driven-dissipative behaviour of the system.
As a result, we get equations for the probabilities ${P_N}\left( t \right)$ (see the Supplemental Material~\ref{sec:SI_2})
\begin{equation}\label{Probabilities_equations_0}
{\frac{\partial {P_0}\left( t \right)}{\partial t}} = {\frac{{d_0}}{{Z_1}}}{P_1}\left( t \right) - {\frac{{p_0}}{{Z_0}}}{P_0}\left( t \right)
\end{equation}
\begin{multline}\label{Probabilities_equations_N}
{\frac{\partial {P_N}\left( t \right)}{\partial t}} = {\frac{{d_N}}{{Z_{N + 1}}}}{P_{N + 1}}\left( t \right) - {\frac{\left( {{d_{N - 1}} + {p_N}} \right)}{{Z_N}}}{P_N}\left( t \right) \\
+ {\frac{{p_{N - 1}}}{{Z_{N - 1}}}}{P_{N - 1}}\left( t \right),\,\,\,\,\text{for}\,\,N>0
\end{multline}
where ${d_N} = \sum\nolimits_{n = 0}^N {{Z_{N - n}}\sum\nolimits_{j = 0}^M {\left( {{\gamma _j} + {\kappa _j}} \right)w_j^{n + 1}} }$ and ${p_N} = \sum\nolimits_{n = 0}^N {{Z_{N - n}}\sum\nolimits_{j = 0}^M {{\kappa _j}w_j^n} }$.

In the case of steady-state non-resonant pumping the stationary solution of Equations~(\ref{Probabilities_equations_0})-(\ref{Probabilities_equations_N}) can be written in recurrent form:
\begin{equation}\label{Probabilities_solution_0}
{P_1} = {\frac{{p_0}}{{d_0}}}{\frac{{Z_1}}{{Z_0}}}{P_0}
\end{equation}
\begin{equation}\label{Probabilities_solution_N}
{P_{N + 1}} = {\frac{\left( {{d_{N - 1}} + {p_N}} \right)}{{d_N}}}{\frac{{Z_{N + 1}}}{{Z_N}}}{P_N} - {\frac{{p_{N - 1}}}{{d_N}}}{\frac{{Z_{N + 1}}}{{Z_{N - 1}}}}{P_{N - 1}}
\end{equation}
On the basis of our theory one can find the complete density matrix of the polaritons at the low polariton branch.

\textit{Formation of non-equilibrium BEC.}
In this section we apply the developed theory to a polariton system consisting of $M + 1$ states equidistant in frequency.
This condition on the frequencies of the polariton states corresponds to a constant density of states in the continuous limit of the two-dimensional system with quadratic dispersion.
The frequency of the $j$-th polariton state is $\Delta {\omega _j} = \left( {{\omega _M} - {\omega _0}} \right) \times j/M$.
We consider the decay rates to be the same for the entire lower polariton branch and equal $\gamma $.
We also assume that the incoherent pumping is characterized by a rate $\kappa $, which does not depend on time, and acts only on the polariton state with $j = M$.

The stationary density matrix of the polariton system under consideration has the form~(\ref{Density_matrix}), where the coefficients  ${P_N}\left( t \right)$ are time-independent given by the expressions~(\ref{Probabilities_solution_0})-(\ref{Probabilities_solution_N}).
Following Eq.~(\ref{Probabilities_solution_0})-(\ref{Probabilities_solution_N}) the stationary density matrix is well-defined by two dimensionless parameters, namely, the ratio of incoherent pumping to dissipation in the states $\kappa /\gamma $ and the ratio $\left( {{\omega _M} - {\omega _0}} \right) /T$.
We analyze how $\kappa $ affects the stationary density matrix of polaritons in the lower branch~(\ref{Density_matrix}).

\begin{figure}
\includegraphics[width=0.93\linewidth]{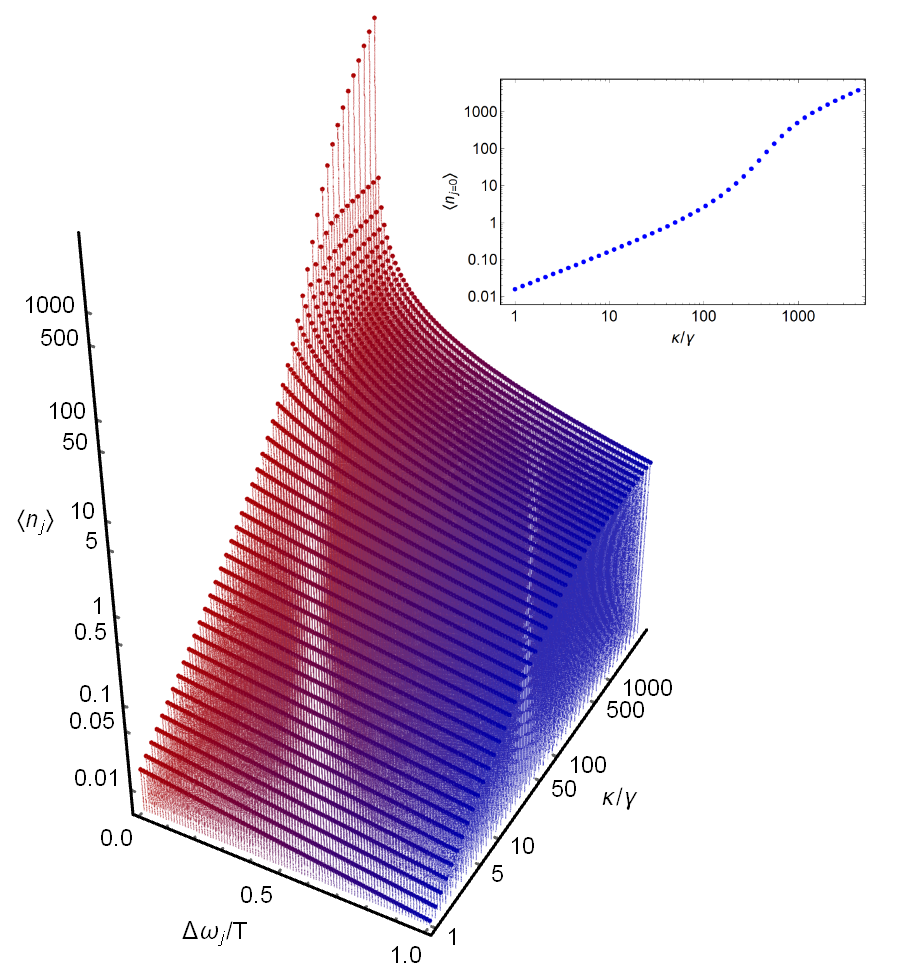}
\caption{\label{fig:Number_of_polaritons} The dependence of the average population of polaritons $\left\langle {{{\hat n}_j}} \right\rangle = \langle \hat a^\dag_j \hat a_j \rangle $ on the incoherent pumping rate $\kappa $ at $M + 1 = 100$, $\left( {{\omega _M} - {\omega _0}} \right)/T = 1$. (Inset) Average population of polaritons in the ground state. }
\end{figure}

The solution for the stationary density matrix~(\ref{Density_matrix}) with the coefficients ${P_N}$, determined by~(\ref{Probabilities_solution_0})-(\ref{Probabilities_solution_N}), allows us to find the average population of the polaritons $\left\langle {{{\hat n}_j}} \right\rangle  = {\mathop{\rm tr}\nolimits} \left( {{{\hat n}_j}\hat \rho } \right)$.
At low pumping rates, the population exponentially decreases with an increase of the state frequency~(Fig.~\ref{fig:Number_of_polaritons}). 
In this case, the exponent does not depend on $\kappa$.
However, for $\kappa$ exceeding a certain threshold value (hereinafter, the condensation threshold), a relatively large number of polaritons accumulates in the ground state~(Fig.~\ref{fig:Number_of_polaritons}). 
This behavior of the average population of polaritons was observed in the experiments~\cite{kasprzak2006bose, balili2007bose, plumhof2014room, sun2017bose} and associated with the formation of a polariton BEC. 
Note that the average polariton density in the ground state $\left\langle {{{\hat n}_{j = 0}}} \right\rangle$ sharply increases at the condensation threshold (Fig.~\ref{fig:Number_of_polaritons}) as has been evidenced in various experiments~\cite{kasprzak2006bose, balili2007bose, plumhof2014room, sun2017bose}. 

\begin{figure}
\includegraphics[width=0.8\linewidth]{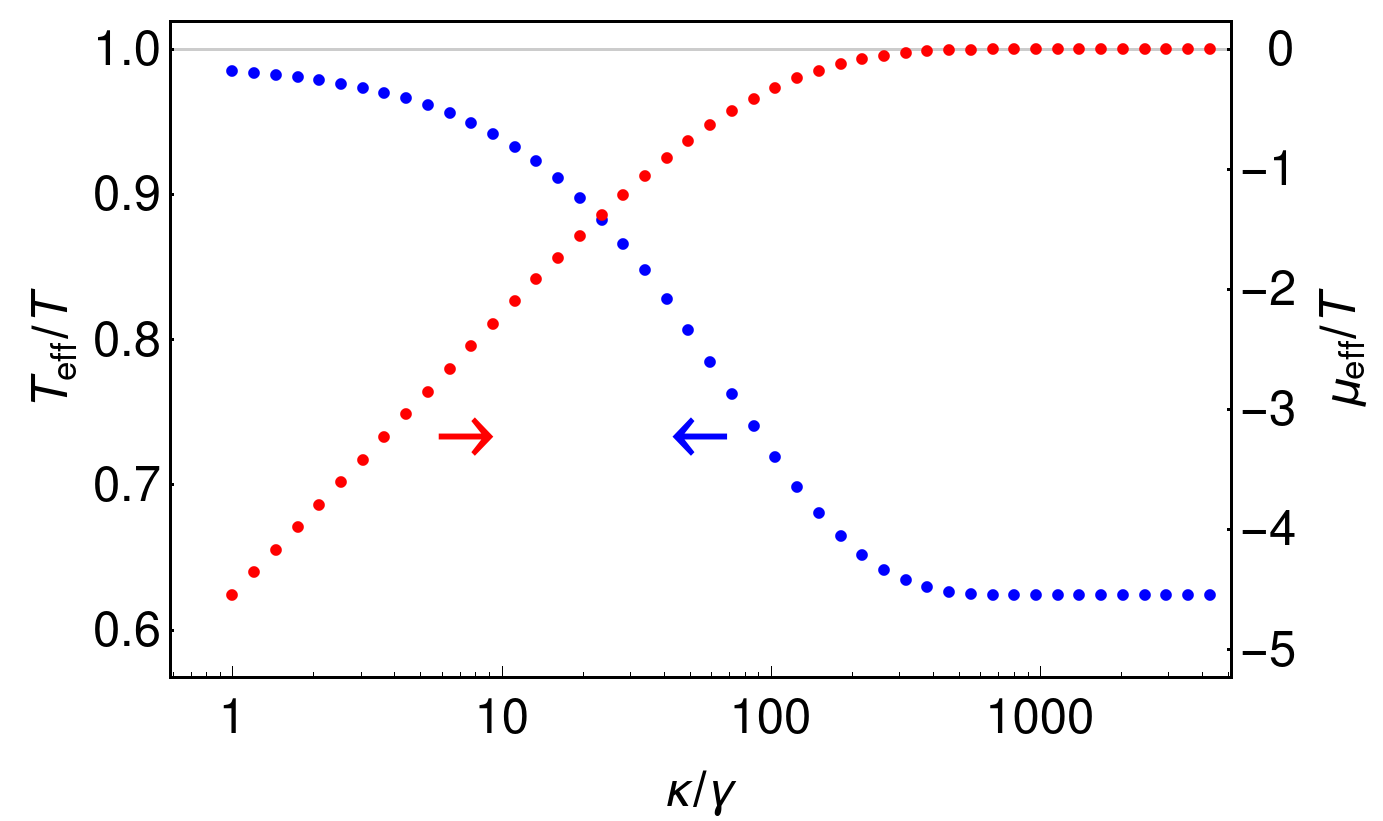}
\caption{\label{fig:Effective_BE} Approximation of the average population of polaritons in non-equilibrium BEC by the equilibrium Bose--Einstein distribution~(\ref{BE}) at $M + 1 = 100$, $\left( {{\omega _M} - {\omega _0}} \right)/T = 1$. (Blue) The ratio of the effective temperature of polaritons to the reservoir temperature as a function of the rate of incoherent pumping. (Red) Effective chemical potential of polaritons as a function of the rate of incoherent pumping.}
\end{figure}

The incoherent pumping drives the system out of  thermodynamic equilibrium.
Therefore, the applicability of equilibrium statistical mechanics approaches to polariton condensates is at least controversial. 
To explore differences and similarities between equilibrium and non-equilibrium BECs, first, we study the average number of polaritons $\left\langle {{{\hat n}_j}} \right\rangle = {\rm{Tr}}\left( {\hat\rho {\hat n_j}} \right) $ and examine the resultant polariton population with respect to Bose--Einstein distribution (see Fig.~\ref{fig:Number_of_polaritons})
\begin{equation}\label{BE}
\left\langle {{{\hat n}_j}} \right\rangle  = \frac{A}{{\exp \left( {{{\left( {{\omega _j} - {\omega _0} - {\mu _{{\rm{eff}}}}} \right)} \mathord{\left/
 {\vphantom {{\left( {{\omega _j} - {\omega _0} - {\mu _{{\rm{eff}}}}} \right)} {{T_{{\rm{eff}}}}}}} \right.
 \kern-\nulldelimiterspace} {{T_{{\rm{eff}}}}}}} \right) - 1}}
\end{equation}
where ${T_{{\rm{eff}}}}$, ${{\mu _{{\rm{eff}}}}}$ and $A$ are extracted by the least squares fit.
For a system in thermodynamic equilibrium  ${T_{{\rm{eff}}}}$ is the equilibrium temperature of the system, ${\mu_{{\rm{eff}}}}$ is the chemical potential, and $A$ is an effective parameter proportional to the number of states around the ground state~\cite{landau2013statistical} defined by size of a system.
Like in thermodynamic equilibrium, we assign ${T_{{\rm{eff}}}}$ to the effective condensate temperature, ${{\mu _{{\rm{eff}}}}}$ to the effective chemical potential of the polariton condensate.
Analogously, we suppose that $A$ does not depend on ${T_{{\rm{eff}}}}$ or ${{\mu _{{\rm{eff}}}}}$. 
Consequently, $A$ should not depend on the rate of incoherent pumping $\kappa$.
The non-equilibrium polariton condensate demonstrates the same trend of decrease ${\mu _{{\rm{eff}}}}$ with increasing $\kappa$~(Fig.~\ref{fig:Effective_BE}), like conventional BECs in equilibrium statistical mechanics.
When the condensation threshold is exceeded, ${\mu _{{\rm{eff}}}}$ becomes almost equal to zero.
However, unlike BEC at thermodynamic equilibrium the non-equilibrium condensate exhibits nontrivial dependence of effective temperature ${T_{{\rm{eff}}}}$ on $\kappa$, as shown in the Figure~\ref{fig:Effective_BE}.
Indeed, ${T_{{\rm{eff}}}}$ collapses below the reservoir temperature~$T$ above condensation threshold leading to stimulated cooling of the polariton gas~(Fig.~\ref{fig:Effective_BE}).

\begin{figure}
\includegraphics[width=0.93\linewidth]{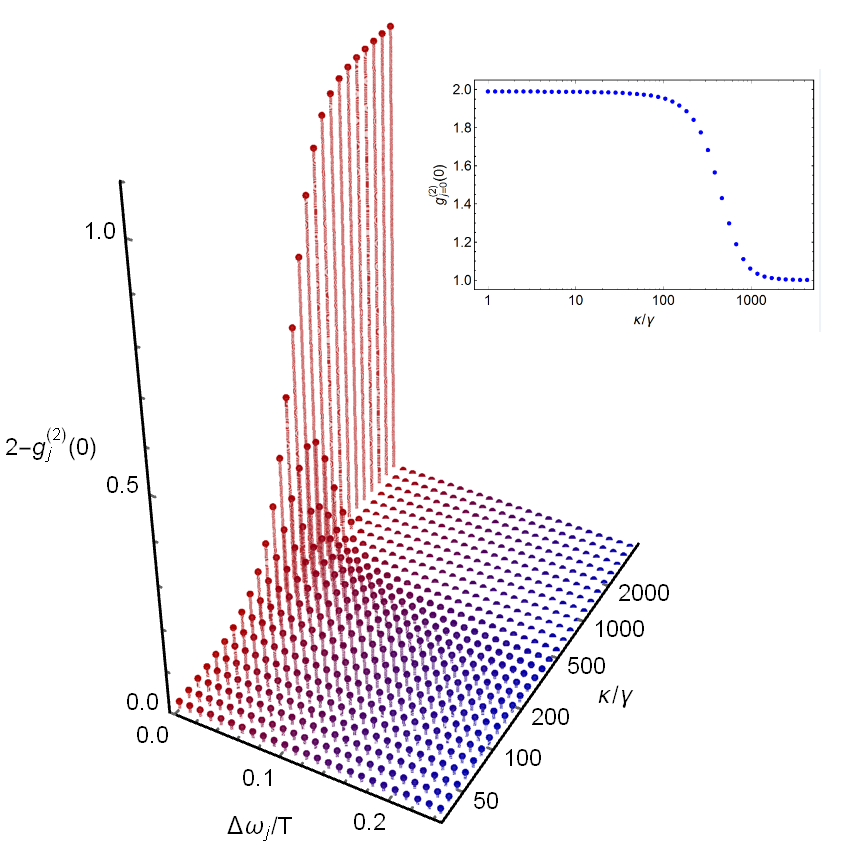}
\caption{\label{fig:Coherence} Correlation properties of non-equilibrium polariton BEC. Second-order autocorrelation function of polaritons in the ground and excited states versus the rate of incoherent pumping at $M + 1 = 100$, $\left( {{\omega _M} - {\omega _0}} \right)/T = 1$. (Inset) Second-order autocorrelation function of the ground state.}
\end{figure}

\textit{Quantum correlations in BEC at the condensation threshold.} 
The main advance enabled by an exact solution~(\ref{Density_matrix}) is that it accesses coherence and high-order correlations of all the modes in polariton BEC. 
Figure~\ref{fig:Coherence} shows the second-order autocorrelation function for polaritons as a function of $\kappa$.
As $\kappa$ increases, two processes take place: on the one hand, the degree of coherence increases, and on the other, the frequency region, in which coherence is formed, narrows.
These two processes lead to a non-monotonic behaviour of the second-order autocorrelation function of the polaritons in the excited states with respect to the rate of incoherent pumping. 
At the same time, the coherence of the ground state monotonically increases. 
Namely, when $\kappa$ exceeds the condensation threshold, the second-order coherence function drops from two to one, see Fig.~\ref{fig:Coherence}. 

\begin{figure}[ht]
\includegraphics[width=0.8\linewidth]{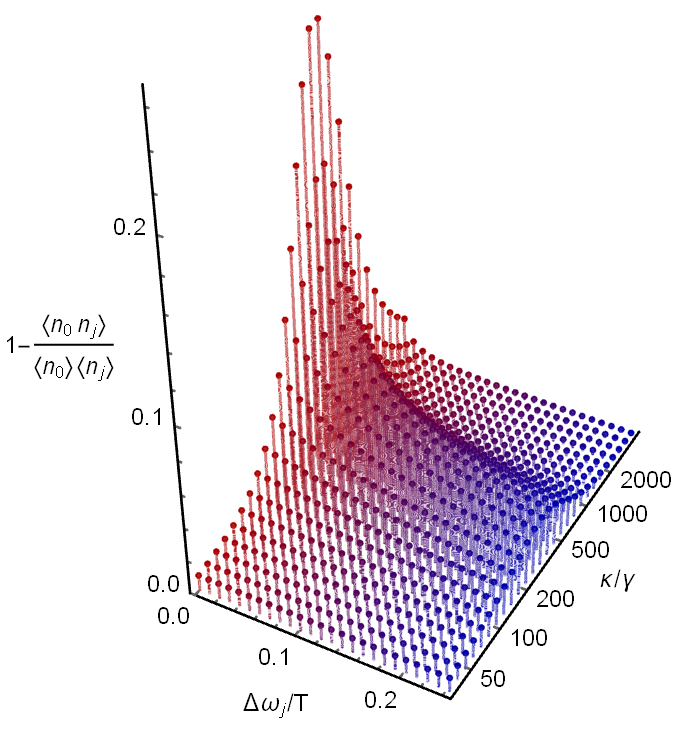}
\caption{\label{fig:nn_nn} Second-order cross-correlation function between the ground polariton state and excited states versus the rate of incoherent pumping at $M + 1 = 100$, $\left( {{\omega _M} - {\omega _0}} \right)/T = 1$.}
\end{figure}

Another peculiar aspect in the formation of polariton BEC is the correlation between different polariton states. 
Although we limit our consideration to cross-correlations between the ground state and other excited states~(Fig.~\ref{fig:nn_nn}), one can extend the analysis on cross-correlation between arbitrary polariton states. 
At the condensation threshold, $\left\langle {{{\hat n}_0}{{\hat n}_{j \ne 0}}} \right\rangle /\left\langle {{{\hat n}_0}} \right\rangle \left\langle {{{\hat n}_{j \ne 0}}} \right\rangle $ is significantly less than unity. 
This means that the ground state and the excited states become correlated in such a way that occupations of ground and excited states at the same time are incompatible events.

\textit{Discussion and conclusion} 
In this letter, we report on an analytical quantum description of non-equilibrium BEC based on solving the master equation for the full polariton density matrix in the limit of fast polariton thermalization. 
While the average polariton density and the effective chemical potential exhibit the usual properties of BEC at thermal equilibrium, we find out the effective temperature strongly deviates from the reservoir temperature, namely polariton gas undergoes noticeable cooling effect.
We also demonstrate that the formation of polariton BEC is accompanied by anti-correlations between the ground polariton state and excited states gradually increasing towards the condensation threshold.

It is worth to mention the developed analytical theory strongly relies on the fast thermalization in bosonic systems.
In fact, dynamics of polariton thermalization depends on the polariton density~\cite{deng2010exciton, kavokin2017microcavities}. Therefore, our approach is valid for sufficiently intense pumping rate nearby the condensate threshold~\cite{kasprzak2006bose, balili2007bose, estrecho2018single, deng2010exciton, wei2019low}. 
However, for long-lived polariton and photon systems thermalization occurs at pumping rates significantly lower condensation threshold~\cite{sun2017bose, weill2019bose}.
Alongside the conventional strongly-coupled microcavities we would like to highlight recently developed plasmonic arrays hybridized with Frenkel excitons as a separate fascinating class of systems demonstrating BEC phenomena at room temperature with extreme thermalization rate on the order of 200~fs~\cite{hakala2018bose, vakevainen2020sub}. 
The ultra-fast thermalization makes our theory especially reliable for such systems in general.

We described polaritons as harmonic oscillators~(\ref{Hamiltonian}) thus neglecting polariton interactions in the system while in some cases, this nonlinearity turns out to be decisive for the formation of condensate~\cite{porras2002polariton, savvidis2000angle}. 
Overall, the theory developed in this letter is applicable when the polariton density is large enough for fast thermalization, but not as high as required for polariton interactions to show up in the substantial energy change of polariton eigenstates. 

For the sake of simplicity, we considered the case when the incoherent pump populates the highest-frequency state only. 
Although in most experiments pumping configuration is slightly different, there are experimental realizations analogous to the considered one~\cite{byrnes2014exciton}.
Nevertheless, in Supplemental Material~\ref{sec:SI_7} we provide results obtained under simultaneous incoherent pump of all considered polariton - the common scenario in experiments. 
The dependencies do not exhibit any significant deviation from the simplified picture considered above.

While the steady-state regime considered in this letter is more illustrative, and the results are easy to interpret, the general equations~(\ref{Probabilities_equations_0})--(\ref{Probabilities_equations_N}) are applicable to the case of pulsed drive as well. 
Moreover, recent experimental advances in quasi-continuous organic polariton condensation~\cite{betzold2019coherence, putintsev2020nano}, bring this theory into the context of room-temperature polariton BEC.

The work was supported by the Russian Science Foundation (Grant No. 20-72-10145). E.S.A thanks the Foundation for the Advancement of Theoretical Physics and Mathematics Basis.

\bibliographystyle{apsrev4-2}
\bibliography{apssamp}

\providecommand{\noopsort}[1]{}\providecommand{\singleletter}[1]{#1}%
\begin{thebibliography}{59}%
\makeatletter
\providecommand \@ifxundefined [1]{%
 \@ifx{#1\undefined}
}%
\providecommand \@ifnum [1]{%
 \ifnum #1\expandafter \@firstoftwo
 \else \expandafter \@secondoftwo
 \fi
}%
\providecommand \@ifx [1]{%
 \ifx #1\expandafter \@firstoftwo
 \else \expandafter \@secondoftwo
 \fi
}%
\providecommand \natexlab [1]{#1}%
\providecommand \enquote  [1]{``#1''}%
\providecommand \bibnamefont  [1]{#1}%
\providecommand \bibfnamefont [1]{#1}%
\providecommand \citenamefont [1]{#1}%
\providecommand \href@noop [0]{\@secondoftwo}%
\providecommand \href [0]{\begingroup \@sanitize@url \@href}%
\providecommand \@href[1]{\@@startlink{#1}\@@href}%
\providecommand \@@href[1]{\endgroup#1\@@endlink}%
\providecommand \@sanitize@url [0]{\catcode `\\12\catcode `\$12\catcode
  `\&12\catcode `\#12\catcode `\^12\catcode `\_12\catcode `\%12\relax}%
\providecommand \@@startlink[1]{}%
\providecommand \@@endlink[0]{}%
\providecommand \url  [0]{\begingroup\@sanitize@url \@url }%
\providecommand \@url [1]{\endgroup\@href {#1}{\urlprefix }}%
\providecommand \urlprefix  [0]{URL }%
\providecommand \Eprint [0]{\href }%
\providecommand \doibase [0]{https://doi.org/}%
\providecommand \selectlanguage [0]{\@gobble}%
\providecommand \bibinfo  [0]{\@secondoftwo}%
\providecommand \bibfield  [0]{\@secondoftwo}%
\providecommand \translation [1]{[#1]}%
\providecommand \BibitemOpen [0]{}%
\providecommand \bibitemStop [0]{}%
\providecommand \bibitemNoStop [0]{.\EOS\space}%
\providecommand \EOS [0]{\spacefactor3000\relax}%
\providecommand \BibitemShut  [1]{\csname bibitem#1\endcsname}%
\let\auto@bib@innerbib\@empty
\bibitem [{\citenamefont {Bose}(1924)}]{bose1924plancks}%
  \BibitemOpen
  \bibfield  {author} {\bibinfo {author} {\bibfnamefont {S.~N.}\ \bibnamefont
  {Bose}},\ }\href@noop {} {\bibfield  {journal} {\bibinfo  {journal}
  {Zeitschrift f{\"u}r Physik}\ }\textbf {\bibinfo {volume} {26}},\ \bibinfo
  {pages} {178} (\bibinfo {year} {1924})}\BibitemShut {NoStop}%
\bibitem [{\citenamefont {Einstein}(1925)}]{einstein1925quantentheorie}%
  \BibitemOpen
  \bibfield  {author} {\bibinfo {author} {\bibfnamefont {A.}~\bibnamefont
  {Einstein}},\ }\href@noop {} {\bibfield  {journal} {\bibinfo  {journal}
  {Sitzungsberichte der Preu{\ss}ischen Akademie der Wissenschaften,
  Physikalisch-Mathematische Klasse, Berlin}\ ,\ \bibinfo {pages} {3}}
  (\bibinfo {year} {1925})}\BibitemShut {NoStop}%
\bibitem [{\citenamefont {Anderson}\ \emph {et~al.}(1995)\citenamefont
  {Anderson}, \citenamefont {Ensher}, \citenamefont {Matthews}, \citenamefont
  {Wieman},\ and\ \citenamefont {Cornell}}]{anderson1995observation}%
  \BibitemOpen
  \bibfield  {author} {\bibinfo {author} {\bibfnamefont {M.~H.}\ \bibnamefont
  {Anderson}}, \bibinfo {author} {\bibfnamefont {J.~R.}\ \bibnamefont
  {Ensher}}, \bibinfo {author} {\bibfnamefont {M.~R.}\ \bibnamefont
  {Matthews}}, \bibinfo {author} {\bibfnamefont {C.~E.}\ \bibnamefont
  {Wieman}},\ and\ \bibinfo {author} {\bibfnamefont {E.~A.}\ \bibnamefont
  {Cornell}},\ }\href@noop {} {\bibfield  {journal} {\bibinfo  {journal}
  {science}\ }\textbf {\bibinfo {volume} {269}},\ \bibinfo {pages} {198}
  (\bibinfo {year} {1995})}\BibitemShut {NoStop}%
\bibitem [{\citenamefont {Davis}\ \emph {et~al.}(1995)\citenamefont {Davis},
  \citenamefont {Mewes}, \citenamefont {Andrews}, \citenamefont {van Druten},
  \citenamefont {Durfee}, \citenamefont {Kurn},\ and\ \citenamefont
  {Ketterle}}]{davis1995bose}%
  \BibitemOpen
  \bibfield  {author} {\bibinfo {author} {\bibfnamefont {K.~B.}\ \bibnamefont
  {Davis}}, \bibinfo {author} {\bibfnamefont {M.-O.}\ \bibnamefont {Mewes}},
  \bibinfo {author} {\bibfnamefont {M.~R.}\ \bibnamefont {Andrews}}, \bibinfo
  {author} {\bibfnamefont {N.~J.}\ \bibnamefont {van Druten}}, \bibinfo
  {author} {\bibfnamefont {D.~S.}\ \bibnamefont {Durfee}}, \bibinfo {author}
  {\bibfnamefont {D.}~\bibnamefont {Kurn}},\ and\ \bibinfo {author}
  {\bibfnamefont {W.}~\bibnamefont {Ketterle}},\ }\href@noop {} {\bibfield
  {journal} {\bibinfo  {journal} {Physical review letters}\ }\textbf {\bibinfo
  {volume} {75}},\ \bibinfo {pages} {3969} (\bibinfo {year}
  {1995})}\BibitemShut {NoStop}%
\bibitem [{\citenamefont {Deng}\ \emph {et~al.}(2003)\citenamefont {Deng},
  \citenamefont {Weihs}, \citenamefont {Snoke}, \citenamefont {Bloch},\ and\
  \citenamefont {Yamamoto}}]{deng2003polariton}%
  \BibitemOpen
  \bibfield  {author} {\bibinfo {author} {\bibfnamefont {H.}~\bibnamefont
  {Deng}}, \bibinfo {author} {\bibfnamefont {G.}~\bibnamefont {Weihs}},
  \bibinfo {author} {\bibfnamefont {D.}~\bibnamefont {Snoke}}, \bibinfo
  {author} {\bibfnamefont {J.}~\bibnamefont {Bloch}},\ and\ \bibinfo {author}
  {\bibfnamefont {Y.}~\bibnamefont {Yamamoto}},\ }\href@noop {} {\bibfield
  {journal} {\bibinfo  {journal} {Proceedings of the National Academy of
  Sciences}\ }\textbf {\bibinfo {volume} {100}},\ \bibinfo {pages} {15318}
  (\bibinfo {year} {2003})}\BibitemShut {NoStop}%
\bibitem [{\citenamefont {Kasprzak}\ \emph {et~al.}(2006)\citenamefont
  {Kasprzak}, \citenamefont {Richard}, \citenamefont {Kundermann},
  \citenamefont {Baas}, \citenamefont {Jeambrun}, \citenamefont {Keeling},
  \citenamefont {Marchetti}, \citenamefont {Szyma{\'n}ska}, \citenamefont
  {Andr{\'e}}, \citenamefont {Staehli} \emph {et~al.}}]{kasprzak2006bose}%
  \BibitemOpen
  \bibfield  {author} {\bibinfo {author} {\bibfnamefont {J.}~\bibnamefont
  {Kasprzak}}, \bibinfo {author} {\bibfnamefont {M.}~\bibnamefont {Richard}},
  \bibinfo {author} {\bibfnamefont {S.}~\bibnamefont {Kundermann}}, \bibinfo
  {author} {\bibfnamefont {A.}~\bibnamefont {Baas}}, \bibinfo {author}
  {\bibfnamefont {P.}~\bibnamefont {Jeambrun}}, \bibinfo {author}
  {\bibfnamefont {J.}~\bibnamefont {Keeling}}, \bibinfo {author} {\bibfnamefont
  {F.}~\bibnamefont {Marchetti}}, \bibinfo {author} {\bibfnamefont
  {M.}~\bibnamefont {Szyma{\'n}ska}}, \bibinfo {author} {\bibfnamefont
  {R.}~\bibnamefont {Andr{\'e}}}, \bibinfo {author} {\bibfnamefont
  {J.}~\bibnamefont {Staehli}}, \emph {et~al.},\ }\href@noop {} {\bibfield
  {journal} {\bibinfo  {journal} {Nature}\ }\textbf {\bibinfo {volume} {443}},\
  \bibinfo {pages} {409} (\bibinfo {year} {2006})}\BibitemShut {NoStop}%
\bibitem [{\citenamefont {Combescot}\ and\ \citenamefont
  {Shiau}(2015)}]{combescot2015excitons}%
  \BibitemOpen
  \bibfield  {author} {\bibinfo {author} {\bibfnamefont {M.}~\bibnamefont
  {Combescot}}\ and\ \bibinfo {author} {\bibfnamefont {S.-Y.}\ \bibnamefont
  {Shiau}},\ }\href@noop {} {\emph {\bibinfo {title} {Excitons and Cooper
  pairs: two composite bosons in many-body physics}}}\ (\bibinfo  {publisher}
  {Oxford University Press},\ \bibinfo {year} {2015})\BibitemShut {NoStop}%
\bibitem [{\citenamefont {Byrnes}\ \emph {et~al.}(2014)\citenamefont {Byrnes},
  \citenamefont {Kim},\ and\ \citenamefont {Yamamoto}}]{byrnes2014exciton}%
  \BibitemOpen
  \bibfield  {author} {\bibinfo {author} {\bibfnamefont {T.}~\bibnamefont
  {Byrnes}}, \bibinfo {author} {\bibfnamefont {N.~Y.}\ \bibnamefont {Kim}},\
  and\ \bibinfo {author} {\bibfnamefont {Y.}~\bibnamefont {Yamamoto}},\
  }\href@noop {} {\bibfield  {journal} {\bibinfo  {journal} {Nature Physics}\
  }\textbf {\bibinfo {volume} {10}},\ \bibinfo {pages} {803} (\bibinfo {year}
  {2014})}\BibitemShut {NoStop}%
\bibitem [{\citenamefont {Wertz}\ \emph {et~al.}(2010)\citenamefont {Wertz},
  \citenamefont {Ferrier}, \citenamefont {Solnyshkov}, \citenamefont {Johne},
  \citenamefont {Sanvitto}, \citenamefont {Lema{\^\i}tre}, \citenamefont
  {Sagnes}, \citenamefont {Grousson}, \citenamefont {Kavokin}, \citenamefont
  {Senellart} \emph {et~al.}}]{wertz2010spontaneous}%
  \BibitemOpen
  \bibfield  {author} {\bibinfo {author} {\bibfnamefont {E.}~\bibnamefont
  {Wertz}}, \bibinfo {author} {\bibfnamefont {L.}~\bibnamefont {Ferrier}},
  \bibinfo {author} {\bibfnamefont {D.}~\bibnamefont {Solnyshkov}}, \bibinfo
  {author} {\bibfnamefont {R.}~\bibnamefont {Johne}}, \bibinfo {author}
  {\bibfnamefont {D.}~\bibnamefont {Sanvitto}}, \bibinfo {author}
  {\bibfnamefont {A.}~\bibnamefont {Lema{\^\i}tre}}, \bibinfo {author}
  {\bibfnamefont {I.}~\bibnamefont {Sagnes}}, \bibinfo {author} {\bibfnamefont
  {R.}~\bibnamefont {Grousson}}, \bibinfo {author} {\bibfnamefont {A.~V.}\
  \bibnamefont {Kavokin}}, \bibinfo {author} {\bibfnamefont {P.}~\bibnamefont
  {Senellart}}, \emph {et~al.},\ }\href@noop {} {\bibfield  {journal} {\bibinfo
   {journal} {Nature physics}\ }\textbf {\bibinfo {volume} {6}},\ \bibinfo
  {pages} {860} (\bibinfo {year} {2010})}\BibitemShut {NoStop}%
\bibitem [{\citenamefont {Balili}\ \emph {et~al.}(2007)\citenamefont {Balili},
  \citenamefont {Hartwell}, \citenamefont {Snoke}, \citenamefont {Pfeiffer},\
  and\ \citenamefont {West}}]{balili2007bose}%
  \BibitemOpen
  \bibfield  {author} {\bibinfo {author} {\bibfnamefont {R.}~\bibnamefont
  {Balili}}, \bibinfo {author} {\bibfnamefont {V.}~\bibnamefont {Hartwell}},
  \bibinfo {author} {\bibfnamefont {D.}~\bibnamefont {Snoke}}, \bibinfo
  {author} {\bibfnamefont {L.}~\bibnamefont {Pfeiffer}},\ and\ \bibinfo
  {author} {\bibfnamefont {K.}~\bibnamefont {West}},\ }\href@noop {} {\bibfield
   {journal} {\bibinfo  {journal} {Science}\ }\textbf {\bibinfo {volume}
  {316}},\ \bibinfo {pages} {1007} (\bibinfo {year} {2007})}\BibitemShut
  {NoStop}%
\bibitem [{\citenamefont {Estrecho}\ \emph {et~al.}(2018)\citenamefont
  {Estrecho}, \citenamefont {Gao}, \citenamefont {Bobrovska}, \citenamefont
  {Fraser}, \citenamefont {Steger}, \citenamefont {Pfeiffer}, \citenamefont
  {West}, \citenamefont {Liew}, \citenamefont {Matuszewski}, \citenamefont
  {Snoke} \emph {et~al.}}]{estrecho2018single}%
  \BibitemOpen
  \bibfield  {author} {\bibinfo {author} {\bibfnamefont {E.}~\bibnamefont
  {Estrecho}}, \bibinfo {author} {\bibfnamefont {T.}~\bibnamefont {Gao}},
  \bibinfo {author} {\bibfnamefont {N.}~\bibnamefont {Bobrovska}}, \bibinfo
  {author} {\bibfnamefont {M.~D.}\ \bibnamefont {Fraser}}, \bibinfo {author}
  {\bibfnamefont {M.}~\bibnamefont {Steger}}, \bibinfo {author} {\bibfnamefont
  {L.}~\bibnamefont {Pfeiffer}}, \bibinfo {author} {\bibfnamefont
  {K.}~\bibnamefont {West}}, \bibinfo {author} {\bibfnamefont {T.~C.~H.}\
  \bibnamefont {Liew}}, \bibinfo {author} {\bibfnamefont {M.}~\bibnamefont
  {Matuszewski}}, \bibinfo {author} {\bibfnamefont {D.~W.}\ \bibnamefont
  {Snoke}}, \emph {et~al.},\ }\href@noop {} {\bibfield  {journal} {\bibinfo
  {journal} {Nature communications}\ }\textbf {\bibinfo {volume} {9}},\
  \bibinfo {pages} {1} (\bibinfo {year} {2018})}\BibitemShut {NoStop}%
\bibitem [{\citenamefont {Klaas}\ \emph {et~al.}(2018)\citenamefont {Klaas},
  \citenamefont {Schlottmann}, \citenamefont {Flayac}, \citenamefont {Laussy},
  \citenamefont {Gericke}, \citenamefont {Schmidt}, \citenamefont {Helversen},
  \citenamefont {Beyer}, \citenamefont {Brodbeck}, \citenamefont {Suchomel}
  \emph {et~al.}}]{klaas2018photon}%
  \BibitemOpen
  \bibfield  {author} {\bibinfo {author} {\bibfnamefont {M.}~\bibnamefont
  {Klaas}}, \bibinfo {author} {\bibfnamefont {E.}~\bibnamefont {Schlottmann}},
  \bibinfo {author} {\bibfnamefont {H.}~\bibnamefont {Flayac}}, \bibinfo
  {author} {\bibfnamefont {F.}~\bibnamefont {Laussy}}, \bibinfo {author}
  {\bibfnamefont {F.}~\bibnamefont {Gericke}}, \bibinfo {author} {\bibfnamefont
  {M.}~\bibnamefont {Schmidt}}, \bibinfo {author} {\bibfnamefont {M.~v.}\
  \bibnamefont {Helversen}}, \bibinfo {author} {\bibfnamefont {J.}~\bibnamefont
  {Beyer}}, \bibinfo {author} {\bibfnamefont {S.}~\bibnamefont {Brodbeck}},
  \bibinfo {author} {\bibfnamefont {H.}~\bibnamefont {Suchomel}}, \emph
  {et~al.},\ }\href@noop {} {\bibfield  {journal} {\bibinfo  {journal}
  {Physical review letters}\ }\textbf {\bibinfo {volume} {121}},\ \bibinfo
  {pages} {047401} (\bibinfo {year} {2018})}\BibitemShut {NoStop}%
\bibitem [{\citenamefont {Sun}\ \emph {et~al.}(2017)\citenamefont {Sun},
  \citenamefont {Wen}, \citenamefont {Yoon}, \citenamefont {Liu}, \citenamefont
  {Steger}, \citenamefont {Pfeiffer}, \citenamefont {West}, \citenamefont
  {Snoke},\ and\ \citenamefont {Nelson}}]{sun2017bose}%
  \BibitemOpen
  \bibfield  {author} {\bibinfo {author} {\bibfnamefont {Y.}~\bibnamefont
  {Sun}}, \bibinfo {author} {\bibfnamefont {P.}~\bibnamefont {Wen}}, \bibinfo
  {author} {\bibfnamefont {Y.}~\bibnamefont {Yoon}}, \bibinfo {author}
  {\bibfnamefont {G.}~\bibnamefont {Liu}}, \bibinfo {author} {\bibfnamefont
  {M.}~\bibnamefont {Steger}}, \bibinfo {author} {\bibfnamefont {L.~N.}\
  \bibnamefont {Pfeiffer}}, \bibinfo {author} {\bibfnamefont {K.}~\bibnamefont
  {West}}, \bibinfo {author} {\bibfnamefont {D.~W.}\ \bibnamefont {Snoke}},\
  and\ \bibinfo {author} {\bibfnamefont {K.~A.}\ \bibnamefont {Nelson}},\
  }\href@noop {} {\bibfield  {journal} {\bibinfo  {journal} {Physical review
  letters}\ }\textbf {\bibinfo {volume} {118}},\ \bibinfo {pages} {016602}
  (\bibinfo {year} {2017})}\BibitemShut {NoStop}%
\bibitem [{\citenamefont {Deng}\ \emph {et~al.}(2010)\citenamefont {Deng},
  \citenamefont {Haug},\ and\ \citenamefont {Yamamoto}}]{deng2010exciton}%
  \BibitemOpen
  \bibfield  {author} {\bibinfo {author} {\bibfnamefont {H.}~\bibnamefont
  {Deng}}, \bibinfo {author} {\bibfnamefont {H.}~\bibnamefont {Haug}},\ and\
  \bibinfo {author} {\bibfnamefont {Y.}~\bibnamefont {Yamamoto}},\ }\href@noop
  {} {\bibfield  {journal} {\bibinfo  {journal} {Reviews of Modern Physics}\
  }\textbf {\bibinfo {volume} {82}},\ \bibinfo {pages} {1489} (\bibinfo {year}
  {2010})}\BibitemShut {NoStop}%
\bibitem [{\citenamefont {Imamoglu}\ \emph {et~al.}(1996)\citenamefont
  {Imamoglu}, \citenamefont {Ram}, \citenamefont {Pau}, \citenamefont
  {Yamamoto} \emph {et~al.}}]{imamog1996nonequilibrium}%
  \BibitemOpen
  \bibfield  {author} {\bibinfo {author} {\bibfnamefont {A.}~\bibnamefont
  {Imamoglu}}, \bibinfo {author} {\bibfnamefont {R.}~\bibnamefont {Ram}},
  \bibinfo {author} {\bibfnamefont {S.}~\bibnamefont {Pau}}, \bibinfo {author}
  {\bibfnamefont {Y.}~\bibnamefont {Yamamoto}}, \emph {et~al.},\ }\href@noop {}
  {\bibfield  {journal} {\bibinfo  {journal} {Physical Review A}\ }\textbf
  {\bibinfo {volume} {53}},\ \bibinfo {pages} {4250} (\bibinfo {year}
  {1996})}\BibitemShut {NoStop}%
\bibitem [{\citenamefont {Keeling}\ and\ \citenamefont
  {K{\'e}na-Cohen}(2020)}]{keeling2020bose}%
  \BibitemOpen
  \bibfield  {author} {\bibinfo {author} {\bibfnamefont {J.}~\bibnamefont
  {Keeling}}\ and\ \bibinfo {author} {\bibfnamefont {S.}~\bibnamefont
  {K{\'e}na-Cohen}},\ }\href@noop {} {\bibfield  {journal} {\bibinfo  {journal}
  {Annual Review of Physical Chemistry}\ }\textbf {\bibinfo {volume} {71}},\
  \bibinfo {pages} {435} (\bibinfo {year} {2020})}\BibitemShut {NoStop}%
\bibitem [{\citenamefont {Wei}\ \emph {et~al.}(2019)\citenamefont {Wei},
  \citenamefont {Rajendran}, \citenamefont {Ohadi}, \citenamefont {Tropf},
  \citenamefont {Gather}, \citenamefont {Turnbull},\ and\ \citenamefont
  {Samuel}}]{wei2019low}%
  \BibitemOpen
  \bibfield  {author} {\bibinfo {author} {\bibfnamefont {M.}~\bibnamefont
  {Wei}}, \bibinfo {author} {\bibfnamefont {S.~K.}\ \bibnamefont {Rajendran}},
  \bibinfo {author} {\bibfnamefont {H.}~\bibnamefont {Ohadi}}, \bibinfo
  {author} {\bibfnamefont {L.}~\bibnamefont {Tropf}}, \bibinfo {author}
  {\bibfnamefont {M.~C.}\ \bibnamefont {Gather}}, \bibinfo {author}
  {\bibfnamefont {G.~A.}\ \bibnamefont {Turnbull}},\ and\ \bibinfo {author}
  {\bibfnamefont {I.~D.}\ \bibnamefont {Samuel}},\ }\href@noop {} {\bibfield
  {journal} {\bibinfo  {journal} {Optica}\ }\textbf {\bibinfo {volume} {6}},\
  \bibinfo {pages} {1124} (\bibinfo {year} {2019})}\BibitemShut {NoStop}%
\bibitem [{\citenamefont {Zasedatelev}\ \emph {et~al.}(2019)\citenamefont
  {Zasedatelev}, \citenamefont {Baranikov}, \citenamefont {Urbonas},
  \citenamefont {Scafirimuto}, \citenamefont {Scherf}, \citenamefont
  {St{\"o}ferle}, \citenamefont {Mahrt},\ and\ \citenamefont
  {Lagoudakis}}]{zasedatelev2019room}%
  \BibitemOpen
  \bibfield  {author} {\bibinfo {author} {\bibfnamefont {A.~V.}\ \bibnamefont
  {Zasedatelev}}, \bibinfo {author} {\bibfnamefont {A.~V.}\ \bibnamefont
  {Baranikov}}, \bibinfo {author} {\bibfnamefont {D.}~\bibnamefont {Urbonas}},
  \bibinfo {author} {\bibfnamefont {F.}~\bibnamefont {Scafirimuto}}, \bibinfo
  {author} {\bibfnamefont {U.}~\bibnamefont {Scherf}}, \bibinfo {author}
  {\bibfnamefont {T.}~\bibnamefont {St{\"o}ferle}}, \bibinfo {author}
  {\bibfnamefont {R.~F.}\ \bibnamefont {Mahrt}},\ and\ \bibinfo {author}
  {\bibfnamefont {P.~G.}\ \bibnamefont {Lagoudakis}},\ }\href@noop {}
  {\bibfield  {journal} {\bibinfo  {journal} {Nature Photonics}\ }\textbf
  {\bibinfo {volume} {13}},\ \bibinfo {pages} {378} (\bibinfo {year}
  {2019})}\BibitemShut {NoStop}%
\bibitem [{\citenamefont {Plumhof}\ \emph {et~al.}(2014)\citenamefont
  {Plumhof}, \citenamefont {St{\"o}ferle}, \citenamefont {Mai}, \citenamefont
  {Scherf},\ and\ \citenamefont {Mahrt}}]{plumhof2014room}%
  \BibitemOpen
  \bibfield  {author} {\bibinfo {author} {\bibfnamefont {J.~D.}\ \bibnamefont
  {Plumhof}}, \bibinfo {author} {\bibfnamefont {T.}~\bibnamefont
  {St{\"o}ferle}}, \bibinfo {author} {\bibfnamefont {L.}~\bibnamefont {Mai}},
  \bibinfo {author} {\bibfnamefont {U.}~\bibnamefont {Scherf}},\ and\ \bibinfo
  {author} {\bibfnamefont {R.~F.}\ \bibnamefont {Mahrt}},\ }\href@noop {}
  {\bibfield  {journal} {\bibinfo  {journal} {Nature materials}\ }\textbf
  {\bibinfo {volume} {13}},\ \bibinfo {pages} {247} (\bibinfo {year}
  {2014})}\BibitemShut {NoStop}%
\bibitem [{\citenamefont {Kavokin}\ and\ \citenamefont
  {Malpuech}(2003)}]{kavokin2003thin}%
  \BibitemOpen
  \bibfield  {author} {\bibinfo {author} {\bibfnamefont {A.}~\bibnamefont
  {Kavokin}}\ and\ \bibinfo {author} {\bibfnamefont {G.}~\bibnamefont
  {Malpuech}},\ }\href@noop {} {\emph {\bibinfo {title} {Thin films and
  nanostructures}}},\ Vol.~\bibinfo {volume} {32}\ (\bibinfo  {publisher}
  {Academic Press},\ \bibinfo {year} {2003})\BibitemShut {NoStop}%
\bibitem [{\citenamefont {Sanvitto}\ and\ \citenamefont
  {K{\'e}na-Cohen}(2016)}]{sanvitto2016road}%
  \BibitemOpen
  \bibfield  {author} {\bibinfo {author} {\bibfnamefont {D.}~\bibnamefont
  {Sanvitto}}\ and\ \bibinfo {author} {\bibfnamefont {S.}~\bibnamefont
  {K{\'e}na-Cohen}},\ }\href@noop {} {\bibfield  {journal} {\bibinfo  {journal}
  {Nature materials}\ }\textbf {\bibinfo {volume} {15}},\ \bibinfo {pages}
  {1061} (\bibinfo {year} {2016})}\BibitemShut {NoStop}%
\bibitem [{\citenamefont {Malpuech}\ \emph {et~al.}(2002)\citenamefont
  {Malpuech}, \citenamefont {Di~Carlo}, \citenamefont {Kavokin}, \citenamefont
  {Baumberg}, \citenamefont {Zamfirescu},\ and\ \citenamefont
  {Lugli}}]{malpuech2002room}%
  \BibitemOpen
  \bibfield  {author} {\bibinfo {author} {\bibfnamefont {G.}~\bibnamefont
  {Malpuech}}, \bibinfo {author} {\bibfnamefont {A.}~\bibnamefont {Di~Carlo}},
  \bibinfo {author} {\bibfnamefont {A.}~\bibnamefont {Kavokin}}, \bibinfo
  {author} {\bibfnamefont {J.~J.}\ \bibnamefont {Baumberg}}, \bibinfo {author}
  {\bibfnamefont {M.}~\bibnamefont {Zamfirescu}},\ and\ \bibinfo {author}
  {\bibfnamefont {P.}~\bibnamefont {Lugli}},\ }\href@noop {} {\bibfield
  {journal} {\bibinfo  {journal} {Applied physics letters}\ }\textbf {\bibinfo
  {volume} {81}},\ \bibinfo {pages} {412} (\bibinfo {year} {2002})}\BibitemShut
  {NoStop}%
\bibitem [{\citenamefont {Kavokin}\ \emph {et~al.}(2017)\citenamefont
  {Kavokin}, \citenamefont {Baumberg}, \citenamefont {Malpuech},\ and\
  \citenamefont {Laussy}}]{kavokin2017microcavities}%
  \BibitemOpen
  \bibfield  {author} {\bibinfo {author} {\bibfnamefont {A.}~\bibnamefont
  {Kavokin}}, \bibinfo {author} {\bibfnamefont {J.~J.}\ \bibnamefont
  {Baumberg}}, \bibinfo {author} {\bibfnamefont {G.}~\bibnamefont {Malpuech}},\
  and\ \bibinfo {author} {\bibfnamefont {F.~P.}\ \bibnamefont {Laussy}},\
  }\href@noop {} {\emph {\bibinfo {title} {Microcavities}}}\ (\bibinfo
  {publisher} {Oxford university press},\ \bibinfo {year} {2017})\BibitemShut
  {NoStop}%
\bibitem [{\citenamefont {Porras}\ \emph {et~al.}(2002)\citenamefont {Porras},
  \citenamefont {Ciuti}, \citenamefont {Baumberg},\ and\ \citenamefont
  {Tejedor}}]{porras2002polariton}%
  \BibitemOpen
  \bibfield  {author} {\bibinfo {author} {\bibfnamefont {D.}~\bibnamefont
  {Porras}}, \bibinfo {author} {\bibfnamefont {C.}~\bibnamefont {Ciuti}},
  \bibinfo {author} {\bibfnamefont {J.}~\bibnamefont {Baumberg}},\ and\
  \bibinfo {author} {\bibfnamefont {C.}~\bibnamefont {Tejedor}},\ }\href@noop
  {} {\bibfield  {journal} {\bibinfo  {journal} {Physical Review B}\ }\textbf
  {\bibinfo {volume} {66}},\ \bibinfo {pages} {085304} (\bibinfo {year}
  {2002})}\BibitemShut {NoStop}%
\bibitem [{\citenamefont {Banyai}\ and\ \citenamefont
  {Gartner}(2002)}]{banyai2002real}%
  \BibitemOpen
  \bibfield  {author} {\bibinfo {author} {\bibfnamefont {L.}~\bibnamefont
  {Banyai}}\ and\ \bibinfo {author} {\bibfnamefont {P.}~\bibnamefont
  {Gartner}},\ }\href@noop {} {\bibfield  {journal} {\bibinfo  {journal}
  {Physical review letters}\ }\textbf {\bibinfo {volume} {88}},\ \bibinfo
  {pages} {210404} (\bibinfo {year} {2002})}\BibitemShut {NoStop}%
\bibitem [{\citenamefont {Cao}\ \emph {et~al.}(2004)\citenamefont {Cao},
  \citenamefont {Doan}, \citenamefont {Thoai},\ and\ \citenamefont
  {Haug}}]{cao2004condensation}%
  \BibitemOpen
  \bibfield  {author} {\bibinfo {author} {\bibfnamefont {H.~T.}\ \bibnamefont
  {Cao}}, \bibinfo {author} {\bibfnamefont {T.}~\bibnamefont {Doan}}, \bibinfo
  {author} {\bibfnamefont {D.~T.}\ \bibnamefont {Thoai}},\ and\ \bibinfo
  {author} {\bibfnamefont {H.}~\bibnamefont {Haug}},\ }\href@noop {} {\bibfield
   {journal} {\bibinfo  {journal} {Physical Review B}\ }\textbf {\bibinfo
  {volume} {69}},\ \bibinfo {pages} {245325} (\bibinfo {year}
  {2004})}\BibitemShut {NoStop}%
\bibitem [{\citenamefont {Doan}\ \emph {et~al.}(2008)\citenamefont {Doan},
  \citenamefont {Cao}, \citenamefont {Thoai},\ and\ \citenamefont
  {Haug}}]{doan2008coherence}%
  \BibitemOpen
  \bibfield  {author} {\bibinfo {author} {\bibfnamefont {T.}~\bibnamefont
  {Doan}}, \bibinfo {author} {\bibfnamefont {H.~T.}\ \bibnamefont {Cao}},
  \bibinfo {author} {\bibfnamefont {D.~T.}\ \bibnamefont {Thoai}},\ and\
  \bibinfo {author} {\bibfnamefont {H.}~\bibnamefont {Haug}},\ }\href@noop {}
  {\bibfield  {journal} {\bibinfo  {journal} {Physical Review B}\ }\textbf
  {\bibinfo {volume} {78}},\ \bibinfo {pages} {205306} (\bibinfo {year}
  {2008})}\BibitemShut {NoStop}%
\bibitem [{\citenamefont {Tassone}\ \emph {et~al.}(1997)\citenamefont
  {Tassone}, \citenamefont {Piermarocchi}, \citenamefont {Savona},
  \citenamefont {Quattropani},\ and\ \citenamefont
  {Schwendimann}}]{tassone1997bottleneck}%
  \BibitemOpen
  \bibfield  {author} {\bibinfo {author} {\bibfnamefont {F.}~\bibnamefont
  {Tassone}}, \bibinfo {author} {\bibfnamefont {C.}~\bibnamefont
  {Piermarocchi}}, \bibinfo {author} {\bibfnamefont {V.}~\bibnamefont
  {Savona}}, \bibinfo {author} {\bibfnamefont {A.}~\bibnamefont
  {Quattropani}},\ and\ \bibinfo {author} {\bibfnamefont {P.}~\bibnamefont
  {Schwendimann}},\ }\href@noop {} {\bibfield  {journal} {\bibinfo  {journal}
  {Physical Review B}\ }\textbf {\bibinfo {volume} {56}},\ \bibinfo {pages}
  {7554} (\bibinfo {year} {1997})}\BibitemShut {NoStop}%
\bibitem [{\citenamefont {Kirton}\ and\ \citenamefont
  {Keeling}(2013)}]{kirton2013nonequilibrium}%
  \BibitemOpen
  \bibfield  {author} {\bibinfo {author} {\bibfnamefont {P.}~\bibnamefont
  {Kirton}}\ and\ \bibinfo {author} {\bibfnamefont {J.}~\bibnamefont
  {Keeling}},\ }\href@noop {} {\bibfield  {journal} {\bibinfo  {journal}
  {Physical review letters}\ }\textbf {\bibinfo {volume} {111}},\ \bibinfo
  {pages} {100404} (\bibinfo {year} {2013})}\BibitemShut {NoStop}%
\bibitem [{\citenamefont {Kirton}\ and\ \citenamefont
  {Keeling}(2015)}]{kirton2015thermalization}%
  \BibitemOpen
  \bibfield  {author} {\bibinfo {author} {\bibfnamefont {P.}~\bibnamefont
  {Kirton}}\ and\ \bibinfo {author} {\bibfnamefont {J.}~\bibnamefont
  {Keeling}},\ }\href@noop {} {\bibfield  {journal} {\bibinfo  {journal}
  {Physical Review A}\ }\textbf {\bibinfo {volume} {91}},\ \bibinfo {pages}
  {033826} (\bibinfo {year} {2015})}\BibitemShut {NoStop}%
\bibitem [{\citenamefont {Strashko}\ \emph {et~al.}(2018)\citenamefont
  {Strashko}, \citenamefont {Kirton},\ and\ \citenamefont
  {Keeling}}]{strashko2018organic}%
  \BibitemOpen
  \bibfield  {author} {\bibinfo {author} {\bibfnamefont {A.}~\bibnamefont
  {Strashko}}, \bibinfo {author} {\bibfnamefont {P.}~\bibnamefont {Kirton}},\
  and\ \bibinfo {author} {\bibfnamefont {J.}~\bibnamefont {Keeling}},\
  }\href@noop {} {\bibfield  {journal} {\bibinfo  {journal} {Physical Review
  Letters}\ }\textbf {\bibinfo {volume} {121}},\ \bibinfo {pages} {193601}
  (\bibinfo {year} {2018})}\BibitemShut {NoStop}%
\bibitem [{\citenamefont {Hartwell}\ and\ \citenamefont
  {Snoke}(2010)}]{hartwell2010numerical}%
  \BibitemOpen
  \bibfield  {author} {\bibinfo {author} {\bibfnamefont {V.}~\bibnamefont
  {Hartwell}}\ and\ \bibinfo {author} {\bibfnamefont {D.}~\bibnamefont
  {Snoke}},\ }\href@noop {} {\bibfield  {journal} {\bibinfo  {journal}
  {Physical Review B}\ }\textbf {\bibinfo {volume} {82}},\ \bibinfo {pages}
  {075307} (\bibinfo {year} {2010})}\BibitemShut {NoStop}%
\bibitem [{\citenamefont {Kasprzak}\ \emph {et~al.}(2008)\citenamefont
  {Kasprzak}, \citenamefont {Solnyshkov}, \citenamefont {Andr{\'e}},
  \citenamefont {Dang},\ and\ \citenamefont
  {Malpuech}}]{kasprzak2008formation}%
  \BibitemOpen
  \bibfield  {author} {\bibinfo {author} {\bibfnamefont {J.}~\bibnamefont
  {Kasprzak}}, \bibinfo {author} {\bibfnamefont {D.}~\bibnamefont
  {Solnyshkov}}, \bibinfo {author} {\bibfnamefont {R.}~\bibnamefont
  {Andr{\'e}}}, \bibinfo {author} {\bibfnamefont {L.~S.}\ \bibnamefont
  {Dang}},\ and\ \bibinfo {author} {\bibfnamefont {G.}~\bibnamefont
  {Malpuech}},\ }\href@noop {} {\bibfield  {journal} {\bibinfo  {journal}
  {Physical review letters}\ }\textbf {\bibinfo {volume} {101}},\ \bibinfo
  {pages} {146404} (\bibinfo {year} {2008})}\BibitemShut {NoStop}%
\bibitem [{\citenamefont {Sanvitto}\ and\ \citenamefont
  {Timofeev}(2012)}]{sanvitto2012exciton}%
  \BibitemOpen
  \bibfield  {author} {\bibinfo {author} {\bibfnamefont {D.}~\bibnamefont
  {Sanvitto}}\ and\ \bibinfo {author} {\bibfnamefont {V.}~\bibnamefont
  {Timofeev}},\ }\href@noop {} {\emph {\bibinfo {title} {Exciton Polaritons in
  Microcavities: New Frontiers}}},\ Vol.\ \bibinfo {volume} {172}\ (\bibinfo
  {publisher} {Springer Science \& Business Media},\ \bibinfo {year}
  {2012})\BibitemShut {NoStop}%
\bibitem [{\citenamefont {Arnardottir}\ \emph {et~al.}(2020)\citenamefont
  {Arnardottir}, \citenamefont {Moilanen}, \citenamefont {Strashko},
  \citenamefont {T{\"o}rm{\"a}},\ and\ \citenamefont
  {Keeling}}]{arnardottir2020multimode}%
  \BibitemOpen
  \bibfield  {author} {\bibinfo {author} {\bibfnamefont {K.~B.}\ \bibnamefont
  {Arnardottir}}, \bibinfo {author} {\bibfnamefont {A.~J.}\ \bibnamefont
  {Moilanen}}, \bibinfo {author} {\bibfnamefont {A.}~\bibnamefont {Strashko}},
  \bibinfo {author} {\bibfnamefont {P.}~\bibnamefont {T{\"o}rm{\"a}}},\ and\
  \bibinfo {author} {\bibfnamefont {J.}~\bibnamefont {Keeling}},\ }\href@noop
  {} {\bibfield  {journal} {\bibinfo  {journal} {Physical Review Letters}\
  }\textbf {\bibinfo {volume} {125}},\ \bibinfo {pages} {233603} (\bibinfo
  {year} {2020})}\BibitemShut {NoStop}%
\bibitem [{\citenamefont {Rubo}\ \emph {et~al.}(2003)\citenamefont {Rubo},
  \citenamefont {Laussy}, \citenamefont {Malpuech}, \citenamefont {Kavokin},\
  and\ \citenamefont {Bigenwald}}]{rubo2003dynamical}%
  \BibitemOpen
  \bibfield  {author} {\bibinfo {author} {\bibfnamefont {Y.~G.}\ \bibnamefont
  {Rubo}}, \bibinfo {author} {\bibfnamefont {F.}~\bibnamefont {Laussy}},
  \bibinfo {author} {\bibfnamefont {G.}~\bibnamefont {Malpuech}}, \bibinfo
  {author} {\bibfnamefont {A.}~\bibnamefont {Kavokin}},\ and\ \bibinfo {author}
  {\bibfnamefont {P.}~\bibnamefont {Bigenwald}},\ }\href@noop {} {\bibfield
  {journal} {\bibinfo  {journal} {Physical review letters}\ }\textbf {\bibinfo
  {volume} {91}},\ \bibinfo {pages} {156403} (\bibinfo {year}
  {2003})}\BibitemShut {NoStop}%
\bibitem [{\citenamefont {Laussy}\ \emph {et~al.}(2004)\citenamefont {Laussy},
  \citenamefont {Malpuech},\ and\ \citenamefont
  {Kavokin}}]{laussy2004spontaneous}%
  \BibitemOpen
  \bibfield  {author} {\bibinfo {author} {\bibfnamefont {F.~P.}\ \bibnamefont
  {Laussy}}, \bibinfo {author} {\bibfnamefont {G.}~\bibnamefont {Malpuech}},\
  and\ \bibinfo {author} {\bibfnamefont {A.}~\bibnamefont {Kavokin}},\
  }\href@noop {} {\bibfield  {journal} {\bibinfo  {journal} {physica status
  solidi (c)}\ }\textbf {\bibinfo {volume} {1}},\ \bibinfo {pages} {1339}
  (\bibinfo {year} {2004})}\BibitemShut {NoStop}%
\bibitem [{\citenamefont {del Valle}\ \emph {et~al.}(2009)\citenamefont {del
  Valle}, \citenamefont {Sanvitto}, \citenamefont {Amo}, \citenamefont
  {Laussy}, \citenamefont {Andr\'e}, \citenamefont {Tejedor},\ and\
  \citenamefont {Vi\~na}}]{PhysRevLett.103.096404}%
  \BibitemOpen
  \bibfield  {author} {\bibinfo {author} {\bibfnamefont {E.}~\bibnamefont {del
  Valle}}, \bibinfo {author} {\bibfnamefont {D.}~\bibnamefont {Sanvitto}},
  \bibinfo {author} {\bibfnamefont {A.}~\bibnamefont {Amo}}, \bibinfo {author}
  {\bibfnamefont {F.~P.}\ \bibnamefont {Laussy}}, \bibinfo {author}
  {\bibfnamefont {R.}~\bibnamefont {Andr\'e}}, \bibinfo {author} {\bibfnamefont
  {C.}~\bibnamefont {Tejedor}},\ and\ \bibinfo {author} {\bibfnamefont
  {L.}~\bibnamefont {Vi\~na}},\ }\href
  {https://doi.org/10.1103/PhysRevLett.103.096404} {\bibfield  {journal}
  {\bibinfo  {journal} {Phys. Rev. Lett.}\ }\textbf {\bibinfo {volume} {103}},\
  \bibinfo {pages} {096404} (\bibinfo {year} {2009})}\BibitemShut {NoStop}%
\bibitem [{\citenamefont
  {Rubo}(2004)}]{https://doi.org/10.1002/pssa.200304065}%
  \BibitemOpen
  \bibfield  {author} {\bibinfo {author} {\bibfnamefont {Y.~G.}\ \bibnamefont
  {Rubo}},\ }\href {https://doi.org/https://doi.org/10.1002/pssa.200304065}
  {\bibfield  {journal} {\bibinfo  {journal} {physica status solidi (a)}\
  }\textbf {\bibinfo {volume} {201}},\ \bibinfo {pages} {641} (\bibinfo {year}
  {2004})}\BibitemShut {NoStop}%
\bibitem [{\citenamefont {Laussy}\ \emph {et~al.}(2003)\citenamefont {Laussy},
  \citenamefont {Rubo}, \citenamefont {Malpuech}, \citenamefont {Kavokin},\
  and\ \citenamefont {Bigenwald}}]{https://doi.org/10.1002/pssc.200303205}%
  \BibitemOpen
  \bibfield  {author} {\bibinfo {author} {\bibfnamefont {F.~P.}\ \bibnamefont
  {Laussy}}, \bibinfo {author} {\bibfnamefont {Y.~G.}\ \bibnamefont {Rubo}},
  \bibinfo {author} {\bibfnamefont {G.}~\bibnamefont {Malpuech}}, \bibinfo
  {author} {\bibfnamefont {A.}~\bibnamefont {Kavokin}},\ and\ \bibinfo {author}
  {\bibfnamefont {P.}~\bibnamefont {Bigenwald}},\ }\href
  {https://doi.org/https://doi.org/10.1002/pssc.200303205} {\bibfield
  {journal} {\bibinfo  {journal} {physica status solidi (c)}\ }\textbf
  {\bibinfo {volume} {n/a}},\ \bibinfo {pages} {1476} (\bibinfo {year}
  {2003})}\BibitemShut {NoStop}%
\bibitem [{\citenamefont {Lagoudakis}\ \emph {et~al.}(2003)\citenamefont
  {Lagoudakis}, \citenamefont {Martin}, \citenamefont {Baumberg}, \citenamefont
  {Qarry}, \citenamefont {Cohen},\ and\ \citenamefont
  {Pfeiffer}}]{lagoudakis2003electron}%
  \BibitemOpen
  \bibfield  {author} {\bibinfo {author} {\bibfnamefont {P.}~\bibnamefont
  {Lagoudakis}}, \bibinfo {author} {\bibfnamefont {M.}~\bibnamefont {Martin}},
  \bibinfo {author} {\bibfnamefont {J.}~\bibnamefont {Baumberg}}, \bibinfo
  {author} {\bibfnamefont {A.}~\bibnamefont {Qarry}}, \bibinfo {author}
  {\bibfnamefont {E.}~\bibnamefont {Cohen}},\ and\ \bibinfo {author}
  {\bibfnamefont {L.}~\bibnamefont {Pfeiffer}},\ }\href@noop {} {\bibfield
  {journal} {\bibinfo  {journal} {Physical review letters}\ }\textbf {\bibinfo
  {volume} {90}},\ \bibinfo {pages} {206401} (\bibinfo {year}
  {2003})}\BibitemShut {NoStop}%
\bibitem [{\citenamefont {Maragkou}\ \emph {et~al.}(2010)\citenamefont
  {Maragkou}, \citenamefont {Grundy}, \citenamefont {Ostatnick{\`y}},\ and\
  \citenamefont {Lagoudakis}}]{maragkou2010longitudinal}%
  \BibitemOpen
  \bibfield  {author} {\bibinfo {author} {\bibfnamefont {M.}~\bibnamefont
  {Maragkou}}, \bibinfo {author} {\bibfnamefont {A.}~\bibnamefont {Grundy}},
  \bibinfo {author} {\bibfnamefont {T.}~\bibnamefont {Ostatnick{\`y}}},\ and\
  \bibinfo {author} {\bibfnamefont {P.}~\bibnamefont {Lagoudakis}},\
  }\href@noop {} {\bibfield  {journal} {\bibinfo  {journal} {Applied Physics
  Letters}\ }\textbf {\bibinfo {volume} {97}},\ \bibinfo {pages} {111110}
  (\bibinfo {year} {2010})}\BibitemShut {NoStop}%
\bibitem [{\citenamefont {Coles}\ \emph {et~al.}(2011)\citenamefont {Coles},
  \citenamefont {Michetti}, \citenamefont {Clark}, \citenamefont {Tsoi},
  \citenamefont {Adawi}, \citenamefont {Kim},\ and\ \citenamefont
  {Lidzey}}]{coles2011vibrationally}%
  \BibitemOpen
  \bibfield  {author} {\bibinfo {author} {\bibfnamefont {D.~M.}\ \bibnamefont
  {Coles}}, \bibinfo {author} {\bibfnamefont {P.}~\bibnamefont {Michetti}},
  \bibinfo {author} {\bibfnamefont {C.}~\bibnamefont {Clark}}, \bibinfo
  {author} {\bibfnamefont {W.~C.}\ \bibnamefont {Tsoi}}, \bibinfo {author}
  {\bibfnamefont {A.~M.}\ \bibnamefont {Adawi}}, \bibinfo {author}
  {\bibfnamefont {J.-S.}\ \bibnamefont {Kim}},\ and\ \bibinfo {author}
  {\bibfnamefont {D.~G.}\ \bibnamefont {Lidzey}},\ }\href@noop {} {\bibfield
  {journal} {\bibinfo  {journal} {Advanced Functional Materials}\ }\textbf
  {\bibinfo {volume} {21}},\ \bibinfo {pages} {3691} (\bibinfo {year}
  {2011})}\BibitemShut {NoStop}%
\bibitem [{\citenamefont {Savvidis}\ \emph {et~al.}(2000)\citenamefont
  {Savvidis}, \citenamefont {Baumberg}, \citenamefont {Stevenson},
  \citenamefont {Skolnick}, \citenamefont {Whittaker},\ and\ \citenamefont
  {Roberts}}]{savvidis2000angle}%
  \BibitemOpen
  \bibfield  {author} {\bibinfo {author} {\bibfnamefont {P.}~\bibnamefont
  {Savvidis}}, \bibinfo {author} {\bibfnamefont {J.}~\bibnamefont {Baumberg}},
  \bibinfo {author} {\bibfnamefont {R.}~\bibnamefont {Stevenson}}, \bibinfo
  {author} {\bibfnamefont {M.}~\bibnamefont {Skolnick}}, \bibinfo {author}
  {\bibfnamefont {D.}~\bibnamefont {Whittaker}},\ and\ \bibinfo {author}
  {\bibfnamefont {J.}~\bibnamefont {Roberts}},\ }\href@noop {} {\bibfield
  {journal} {\bibinfo  {journal} {Physical review letters}\ }\textbf {\bibinfo
  {volume} {84}},\ \bibinfo {pages} {1547} (\bibinfo {year}
  {2000})}\BibitemShut {NoStop}%
\bibitem [{\citenamefont {Breuer}\ \emph {et~al.}(2002)\citenamefont {Breuer},
  \citenamefont {Petruccione} \emph {et~al.}}]{breuer2002theory}%
  \BibitemOpen
  \bibfield  {author} {\bibinfo {author} {\bibfnamefont {H.-P.}\ \bibnamefont
  {Breuer}}, \bibinfo {author} {\bibfnamefont {F.}~\bibnamefont {Petruccione}},
  \emph {et~al.},\ }\href@noop {} {\emph {\bibinfo {title} {The theory of open
  quantum systems}}}\ (\bibinfo  {publisher} {Oxford University Press on
  Demand},\ \bibinfo {year} {2002})\BibitemShut {NoStop}%
\bibitem [{\citenamefont {Scully}\ and\ \citenamefont
  {Zubairy}(1999)}]{scully1999quantum}%
  \BibitemOpen
  \bibfield  {author} {\bibinfo {author} {\bibfnamefont {M.~O.}\ \bibnamefont
  {Scully}}\ and\ \bibinfo {author} {\bibfnamefont {M.~S.}\ \bibnamefont
  {Zubairy}},\ }\href@noop {} {\bibinfo {title} {Quantum optics}} (\bibinfo
  {year} {1999})\BibitemShut {NoStop}%
\bibitem [{\citenamefont {Litinskaya}\ \emph {et~al.}(2004)\citenamefont
  {Litinskaya}, \citenamefont {Reineker},\ and\ \citenamefont
  {Agranovich}}]{litinskaya2004fast}%
  \BibitemOpen
  \bibfield  {author} {\bibinfo {author} {\bibfnamefont {M.}~\bibnamefont
  {Litinskaya}}, \bibinfo {author} {\bibfnamefont {P.}~\bibnamefont
  {Reineker}},\ and\ \bibinfo {author} {\bibfnamefont {V.}~\bibnamefont
  {Agranovich}},\ }\href@noop {} {\bibfield  {journal} {\bibinfo  {journal}
  {Journal of luminescence}\ }\textbf {\bibinfo {volume} {110}},\ \bibinfo
  {pages} {364} (\bibinfo {year} {2004})}\BibitemShut {NoStop}%
\bibitem [{\citenamefont {Mazza}\ \emph {et~al.}(2009)\citenamefont {Mazza},
  \citenamefont {Fontanesi},\ and\ \citenamefont
  {La~Rocca}}]{mazza2009organic}%
  \BibitemOpen
  \bibfield  {author} {\bibinfo {author} {\bibfnamefont {L.}~\bibnamefont
  {Mazza}}, \bibinfo {author} {\bibfnamefont {L.}~\bibnamefont {Fontanesi}},\
  and\ \bibinfo {author} {\bibfnamefont {G.}~\bibnamefont {La~Rocca}},\
  }\href@noop {} {\bibfield  {journal} {\bibinfo  {journal} {Physical Review
  B}\ }\textbf {\bibinfo {volume} {80}},\ \bibinfo {pages} {235314} (\bibinfo
  {year} {2009})}\BibitemShut {NoStop}%
\bibitem [{\citenamefont {Bittner}\ and\ \citenamefont
  {Silva}(2012)}]{bittner2012estimating}%
  \BibitemOpen
  \bibfield  {author} {\bibinfo {author} {\bibfnamefont {E.~R.}\ \bibnamefont
  {Bittner}}\ and\ \bibinfo {author} {\bibfnamefont {C.}~\bibnamefont
  {Silva}},\ }\href@noop {} {\bibfield  {journal} {\bibinfo  {journal} {The
  Journal of chemical physics}\ }\textbf {\bibinfo {volume} {136}},\ \bibinfo
  {pages} {034510} (\bibinfo {year} {2012})}\BibitemShut {NoStop}%
\bibitem [{\citenamefont {{\'C}wik}\ \emph {et~al.}(2014)\citenamefont
  {{\'C}wik}, \citenamefont {Reja}, \citenamefont {Littlewood},\ and\
  \citenamefont {Keeling}}]{cwik2014polariton}%
  \BibitemOpen
  \bibfield  {author} {\bibinfo {author} {\bibfnamefont {J.~A.}\ \bibnamefont
  {{\'C}wik}}, \bibinfo {author} {\bibfnamefont {S.}~\bibnamefont {Reja}},
  \bibinfo {author} {\bibfnamefont {P.~B.}\ \bibnamefont {Littlewood}},\ and\
  \bibinfo {author} {\bibfnamefont {J.}~\bibnamefont {Keeling}},\ }\href@noop
  {} {\bibfield  {journal} {\bibinfo  {journal} {EPL (Europhysics Letters)}\
  }\textbf {\bibinfo {volume} {105}},\ \bibinfo {pages} {47009} (\bibinfo
  {year} {2014})}\BibitemShut {NoStop}%
\bibitem [{\citenamefont {Somaschi}\ \emph {et~al.}(2011)\citenamefont
  {Somaschi}, \citenamefont {Mouchliadis}, \citenamefont {Coles}, \citenamefont
  {Perakis}, \citenamefont {Lidzey}, \citenamefont {Lagoudakis},\ and\
  \citenamefont {Savvidis}}]{somaschi2011ultrafast}%
  \BibitemOpen
  \bibfield  {author} {\bibinfo {author} {\bibfnamefont {N.}~\bibnamefont
  {Somaschi}}, \bibinfo {author} {\bibfnamefont {L.}~\bibnamefont
  {Mouchliadis}}, \bibinfo {author} {\bibfnamefont {D.}~\bibnamefont {Coles}},
  \bibinfo {author} {\bibfnamefont {I.}~\bibnamefont {Perakis}}, \bibinfo
  {author} {\bibfnamefont {D.}~\bibnamefont {Lidzey}}, \bibinfo {author}
  {\bibfnamefont {P.}~\bibnamefont {Lagoudakis}},\ and\ \bibinfo {author}
  {\bibfnamefont {P.}~\bibnamefont {Savvidis}},\ }\href@noop {} {\bibfield
  {journal} {\bibinfo  {journal} {Applied Physics Letters}\ }\textbf {\bibinfo
  {volume} {99}},\ \bibinfo {pages} {209} (\bibinfo {year} {2011})}\BibitemShut
  {NoStop}%
\bibitem [{\citenamefont {Ramezani}\ \emph {et~al.}(2018)\citenamefont
  {Ramezani}, \citenamefont {Le-Van}, \citenamefont {Halpin},\ and\
  \citenamefont {Rivas}}]{ramezani2018nonlinear}%
  \BibitemOpen
  \bibfield  {author} {\bibinfo {author} {\bibfnamefont {M.}~\bibnamefont
  {Ramezani}}, \bibinfo {author} {\bibfnamefont {Q.}~\bibnamefont {Le-Van}},
  \bibinfo {author} {\bibfnamefont {A.}~\bibnamefont {Halpin}},\ and\ \bibinfo
  {author} {\bibfnamefont {J.~G.}\ \bibnamefont {Rivas}},\ }\href@noop {}
  {\bibfield  {journal} {\bibinfo  {journal} {Physical Review Letters}\
  }\textbf {\bibinfo {volume} {121}},\ \bibinfo {pages} {243904} (\bibinfo
  {year} {2018})}\BibitemShut {NoStop}%
\bibitem [{\citenamefont {Shishkov}\ \emph {et~al.}(2018)\citenamefont
  {Shishkov}, \citenamefont {Andrianov}, \citenamefont {Pukhov}, \citenamefont
  {Vinogradov},\ and\ \citenamefont {Lisyansky}}]{shishkov2018zeroth}%
  \BibitemOpen
  \bibfield  {author} {\bibinfo {author} {\bibfnamefont {V.~Y.}\ \bibnamefont
  {Shishkov}}, \bibinfo {author} {\bibfnamefont {E.}~\bibnamefont {Andrianov}},
  \bibinfo {author} {\bibfnamefont {A.}~\bibnamefont {Pukhov}}, \bibinfo
  {author} {\bibfnamefont {A.}~\bibnamefont {Vinogradov}},\ and\ \bibinfo
  {author} {\bibfnamefont {A.}~\bibnamefont {Lisyansky}},\ }\href@noop {}
  {\bibfield  {journal} {\bibinfo  {journal} {Physical Review E}\ }\textbf
  {\bibinfo {volume} {98}},\ \bibinfo {pages} {022132} (\bibinfo {year}
  {2018})}\BibitemShut {NoStop}%
\bibitem [{\citenamefont {Landau}\ and\ \citenamefont
  {Lifshitz}(2013)}]{landau2013statistical}%
  \BibitemOpen
  \bibfield  {author} {\bibinfo {author} {\bibfnamefont {L.~D.}\ \bibnamefont
  {Landau}}\ and\ \bibinfo {author} {\bibfnamefont {E.~M.}\ \bibnamefont
  {Lifshitz}},\ }\href@noop {} {\emph {\bibinfo {title} {Statistical Physics:
  Volume 5}}},\ Vol.~\bibinfo {volume} {5}\ (\bibinfo  {publisher} {Elsevier},\
  \bibinfo {year} {2013})\BibitemShut {NoStop}%
\bibitem [{\citenamefont {Weill}\ \emph {et~al.}(2019)\citenamefont {Weill},
  \citenamefont {Bekker}, \citenamefont {Levit},\ and\ \citenamefont
  {Fischer}}]{weill2019bose}%
  \BibitemOpen
  \bibfield  {author} {\bibinfo {author} {\bibfnamefont {R.}~\bibnamefont
  {Weill}}, \bibinfo {author} {\bibfnamefont {A.}~\bibnamefont {Bekker}},
  \bibinfo {author} {\bibfnamefont {B.}~\bibnamefont {Levit}},\ and\ \bibinfo
  {author} {\bibfnamefont {B.}~\bibnamefont {Fischer}},\ }\href@noop {}
  {\bibfield  {journal} {\bibinfo  {journal} {Nature communications}\ }\textbf
  {\bibinfo {volume} {10}},\ \bibinfo {pages} {1} (\bibinfo {year}
  {2019})}\BibitemShut {NoStop}%
\bibitem [{\citenamefont {Hakala}\ \emph {et~al.}(2018)\citenamefont {Hakala},
  \citenamefont {Moilanen}, \citenamefont {V{\"a}kev{\"a}inen}, \citenamefont
  {Guo}, \citenamefont {Martikainen}, \citenamefont {Daskalakis}, \citenamefont
  {Rekola}, \citenamefont {Julku},\ and\ \citenamefont
  {T{\"o}rm{\"a}}}]{hakala2018bose}%
  \BibitemOpen
  \bibfield  {author} {\bibinfo {author} {\bibfnamefont {T.~K.}\ \bibnamefont
  {Hakala}}, \bibinfo {author} {\bibfnamefont {A.~J.}\ \bibnamefont
  {Moilanen}}, \bibinfo {author} {\bibfnamefont {A.~I.}\ \bibnamefont
  {V{\"a}kev{\"a}inen}}, \bibinfo {author} {\bibfnamefont {R.}~\bibnamefont
  {Guo}}, \bibinfo {author} {\bibfnamefont {J.-P.}\ \bibnamefont
  {Martikainen}}, \bibinfo {author} {\bibfnamefont {K.~S.}\ \bibnamefont
  {Daskalakis}}, \bibinfo {author} {\bibfnamefont {H.~T.}\ \bibnamefont
  {Rekola}}, \bibinfo {author} {\bibfnamefont {A.}~\bibnamefont {Julku}},\ and\
  \bibinfo {author} {\bibfnamefont {P.}~\bibnamefont {T{\"o}rm{\"a}}},\
  }\href@noop {} {\bibfield  {journal} {\bibinfo  {journal} {Nature Physics}\
  }\textbf {\bibinfo {volume} {14}},\ \bibinfo {pages} {739} (\bibinfo {year}
  {2018})}\BibitemShut {NoStop}%
\bibitem [{\citenamefont {V{\"a}kev{\"a}inen}\ \emph
  {et~al.}(2020)\citenamefont {V{\"a}kev{\"a}inen}, \citenamefont {Moilanen},
  \citenamefont {Ne{\v{c}}ada}, \citenamefont {Hakala}, \citenamefont
  {Daskalakis},\ and\ \citenamefont {T{\"o}rm{\"a}}}]{vakevainen2020sub}%
  \BibitemOpen
  \bibfield  {author} {\bibinfo {author} {\bibfnamefont {A.~I.}\ \bibnamefont
  {V{\"a}kev{\"a}inen}}, \bibinfo {author} {\bibfnamefont {A.~J.}\ \bibnamefont
  {Moilanen}}, \bibinfo {author} {\bibfnamefont {M.}~\bibnamefont
  {Ne{\v{c}}ada}}, \bibinfo {author} {\bibfnamefont {T.~K.}\ \bibnamefont
  {Hakala}}, \bibinfo {author} {\bibfnamefont {K.~S.}\ \bibnamefont
  {Daskalakis}},\ and\ \bibinfo {author} {\bibfnamefont {P.}~\bibnamefont
  {T{\"o}rm{\"a}}},\ }\href@noop {} {\bibfield  {journal} {\bibinfo  {journal}
  {Nature communications}\ }\textbf {\bibinfo {volume} {11}},\ \bibinfo {pages}
  {1} (\bibinfo {year} {2020})}\BibitemShut {NoStop}%
\bibitem [{\citenamefont {Betzold}\ \emph {et~al.}(2019)\citenamefont
  {Betzold}, \citenamefont {Dusel}, \citenamefont {Kyriienko}, \citenamefont
  {Dietrich}, \citenamefont {Klembt}, \citenamefont {Ohmer}, \citenamefont
  {Fischer}, \citenamefont {Shelykh}, \citenamefont {Schneider},\ and\
  \citenamefont {H\"{o}fling}}]{betzold2019coherence}%
  \BibitemOpen
  \bibfield  {author} {\bibinfo {author} {\bibfnamefont {S.}~\bibnamefont
  {Betzold}}, \bibinfo {author} {\bibfnamefont {M.}~\bibnamefont {Dusel}},
  \bibinfo {author} {\bibfnamefont {O.}~\bibnamefont {Kyriienko}}, \bibinfo
  {author} {\bibfnamefont {C.~P.}\ \bibnamefont {Dietrich}}, \bibinfo {author}
  {\bibfnamefont {S.}~\bibnamefont {Klembt}}, \bibinfo {author} {\bibfnamefont
  {J.}~\bibnamefont {Ohmer}}, \bibinfo {author} {\bibfnamefont
  {U.}~\bibnamefont {Fischer}}, \bibinfo {author} {\bibfnamefont {I.~A.}\
  \bibnamefont {Shelykh}}, \bibinfo {author} {\bibfnamefont {C.}~\bibnamefont
  {Schneider}},\ and\ \bibinfo {author} {\bibfnamefont {S.}~\bibnamefont
  {H\"{o}fling}},\ }\href@noop {} {\bibfield  {journal} {\bibinfo  {journal}
  {ACS Photonics}\ }\textbf {\bibinfo {volume} {7}},\ \bibinfo {pages} {384}
  (\bibinfo {year} {2019})}\BibitemShut {NoStop}%
\bibitem [{\citenamefont {Putintsev}\ \emph {et~al.}(2020)\citenamefont
  {Putintsev}, \citenamefont {Zasedatelev}, \citenamefont {McGhee},
  \citenamefont {Cookson}, \citenamefont {Georgiou}, \citenamefont {Sannikov},
  \citenamefont {Lidzey},\ and\ \citenamefont
  {Lagoudakis}}]{putintsev2020nano}%
  \BibitemOpen
  \bibfield  {author} {\bibinfo {author} {\bibfnamefont {A.}~\bibnamefont
  {Putintsev}}, \bibinfo {author} {\bibfnamefont {A.}~\bibnamefont
  {Zasedatelev}}, \bibinfo {author} {\bibfnamefont {K.~E.}\ \bibnamefont
  {McGhee}}, \bibinfo {author} {\bibfnamefont {T.}~\bibnamefont {Cookson}},
  \bibinfo {author} {\bibfnamefont {K.}~\bibnamefont {Georgiou}}, \bibinfo
  {author} {\bibfnamefont {D.}~\bibnamefont {Sannikov}}, \bibinfo {author}
  {\bibfnamefont {D.~G.}\ \bibnamefont {Lidzey}},\ and\ \bibinfo {author}
  {\bibfnamefont {P.}~\bibnamefont {Lagoudakis}},\ }\href@noop {} {\bibfield
  {journal} {\bibinfo  {journal} {Applied Physics Letters}\ }\textbf {\bibinfo
  {volume} {117}},\ \bibinfo {pages} {123302} (\bibinfo {year}
  {2020})}\BibitemShut {NoStop}%
\end{thebibliography}%

\newpage

\begin{widetext}
\section{\label{sec:SI}SUPPLEMENTAL MATERIAL}

\subsection{\label{sec:SI_1} Properties of the partition function ${Z_N}$}
Some important properties of the partition function ${Z_N}$, defined by~(\ref{Stat_sum}), can be obtained using its generating function 
\begin{equation} \label{Stat_sum_func}
f\left( x \right) = \sum\limits_{N = 0}^{ + \infty } {{x^N}{Z_N}}  = \prod\limits_{j = 0}^M {\frac{1}{1 - {w_j}x}},
\end{equation}
where $M + 1$ is the total number of polariton states, ${w_j} = \exp \left( { - \hbar \left( {{\omega _j} - {\omega _0}} \right)/T} \right)$, ${\omega _j}$ is the frequency of the $j$th state and $T$ is the reservoir temperature. 
The generating function $f\left( x \right)$ is defined for $0 \le x < 1$.

Using the generating function~(\ref{Stat_sum_func}), one can obtain the following properties of~${Z_N}$
\begin{equation} \label{Stat_sum_inf}
\mathop {\lim }\limits_{N \to  + \infty } {Z_N} = \mathop {\lim }\limits_{x \to 1 - 0} \left( {1 - x} \right)f\left( x \right)
\end{equation}
\begin{equation} \label{Stat_sum_diff}
w_j\frac{\partial {Z_N}}{\partial {w_j}} = \sum\limits_{n = 0}^N {w_j^{n}{Z_{N - n}}} - Z_{N} 
\end{equation}

\subsection{\label{sec:SI_2}Derivation of Eqs.~(\ref{Probabilities_equations_0})-(\ref{Probabilities_equations_N})}
Substitution of the general expression for the density matrix in the limit of fast thermalization~(\ref{Density_matrix}) into the master equation~(\ref{Master_equation}) leads to 
\begin{multline} \label{For_probabilities_equations}
\sum\limits_{N = 0}^{ + \infty } {\frac{\partial {P_N}\left( t \right)}{\partial t}}{\frac{1}{{Z_N}}}
\sum\limits_{{n_0} + ... + {n_M} = N} {w_0^{{n_0}}...w_M^{{n_M}}{{\hat r}_{{n_0}...{n_M}}}} =\\
= \sum\limits_{N = 0}^{ + \infty } \frac{{P_N}\left( t \right)}{Z_N} 
\sum\limits_{{n_0} + ... + {n_M} = N} {w_0^{{n_0}}...w_M^{{n_M}}{{ L}_{{\rm{diss}}}}\left( {{{\hat r}_{{n_0}...{n_M}}}} \right)}    
+\sum\limits_{N = 0}^{ + \infty } \frac{{P_N}\left( t \right)}{Z_N} 
\sum\limits_{{n_0} + ... + {n_M} = N} {w_0^{{n_0}}...w_M^{{n_M}}{{L}_{{\rm{pump}}}}\left( {{{\hat r}_{{n_0}...{n_M}}}} \right)}  
\end{multline}
where ${\hat r_{{n_0}...{n_M}}} = \left| {{n_0},...,{n_M}} \right\rangle \left\langle {{n_0},...,{n_M}} \right|$. To obtain Equations~(\ref{Probabilities_equations_0})-(\ref{Probabilities_equations_N}) it is necessary to project Equation~(\ref{For_probabilities_equations}) to states with the total number of polaritons $N$.

\subsection{\label{sec:SI_4}Density matrix of the polaritons in the ground state }
Figure~\ref{fig:Zero_density_matrix} shows the diagonal elements of the polariton density matrix $\rho _{N,N}^{{\rm{condensate}}}$ for $j = 0$, defined by ${\hat \rho _0} = \sum\nolimits_N {\left| N \right\rangle \left\langle N \right|\rho _{N,N}^{{\rm{condensate}}}} $, where ${\hat \rho _0} = {{\mathop{\rm tr}\nolimits} _1}{{\mathop{\rm tr}\nolimits} _2}...{{\mathop{\rm tr}\nolimits} _{M}}\left( {\hat \rho } \right)$, $\hat \rho $ is the full density matrix of the lower polaritons and ${{\mathop{\rm tr}\nolimits} _j}$ is partial trace over the $j$th state.
As one can see from Figure~\ref{fig:Zero_density_matrix}, at pumping rates below the threshold, the maximal diagonal element of the density matrix is $\rho _{0,0}^{{\rm{condensate}}}$. 
However, at pumping rates above the threshold, the maximal diagonal element of the density matrix $\rho _{N,N}^{{\rm{condensate}}}$ corresponds to $N > 0$, and the shape of the distribution of the diagonal elements of the density matrix is getting close to a coherent distribution.

\begin{figure}[ht]
\includegraphics[width=0.4\linewidth]{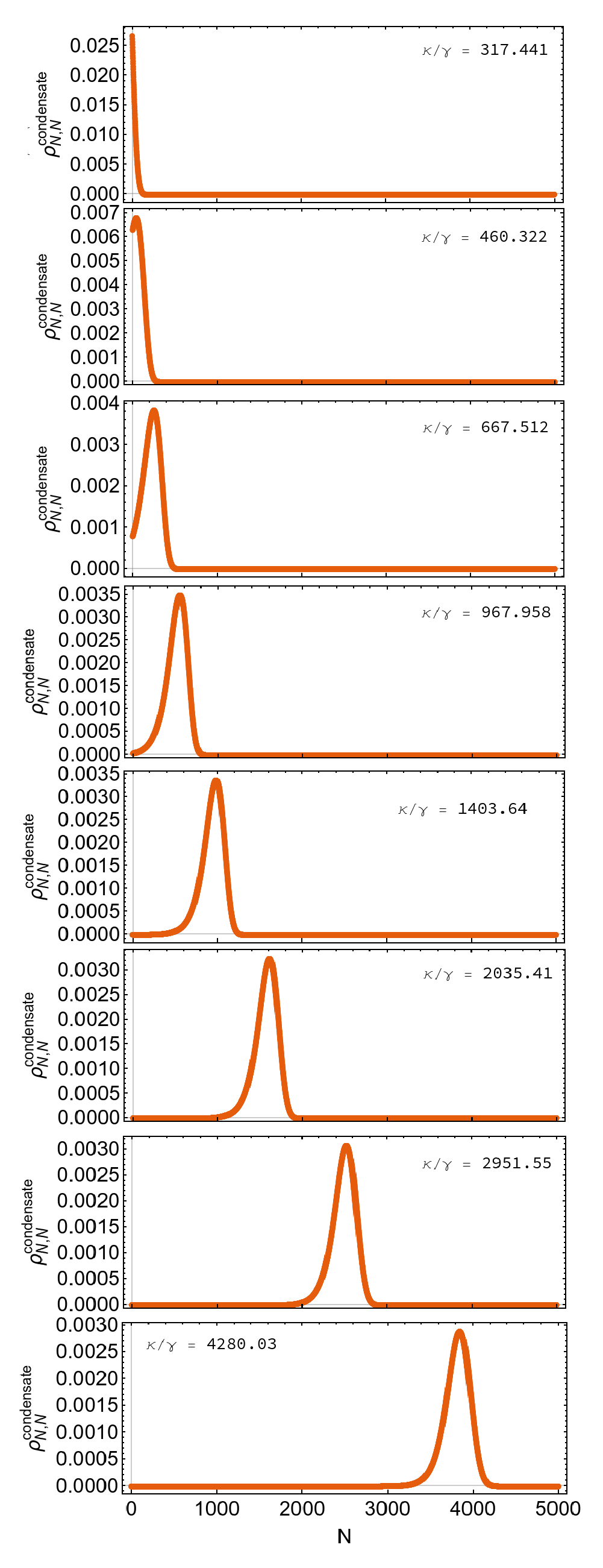}
\caption{\label{fig:Zero_density_matrix} Formation of BEC. Diagonal elements of the density matrix of the ground state at different rates of incoherent pumping.}
\end{figure}

\subsection{\label{sec:SI_5}Approximation of the average population of polaritons by the Bose--Einstein distribution}
Figure~\ref{fig:BE_approximation} shows the calculated distribution of the average population of polaritons and its approximation by the Bose--Einstein distribution. 
Figure~\ref{fig:BE_approximation} shows the relative error arising from this approximation. 
One can see that the Bose--Einstein distribution approximates quite well the stationary distribution of the non-equilibrium polariton BEC.

\begin{figure}
\includegraphics[width=0.5\linewidth]{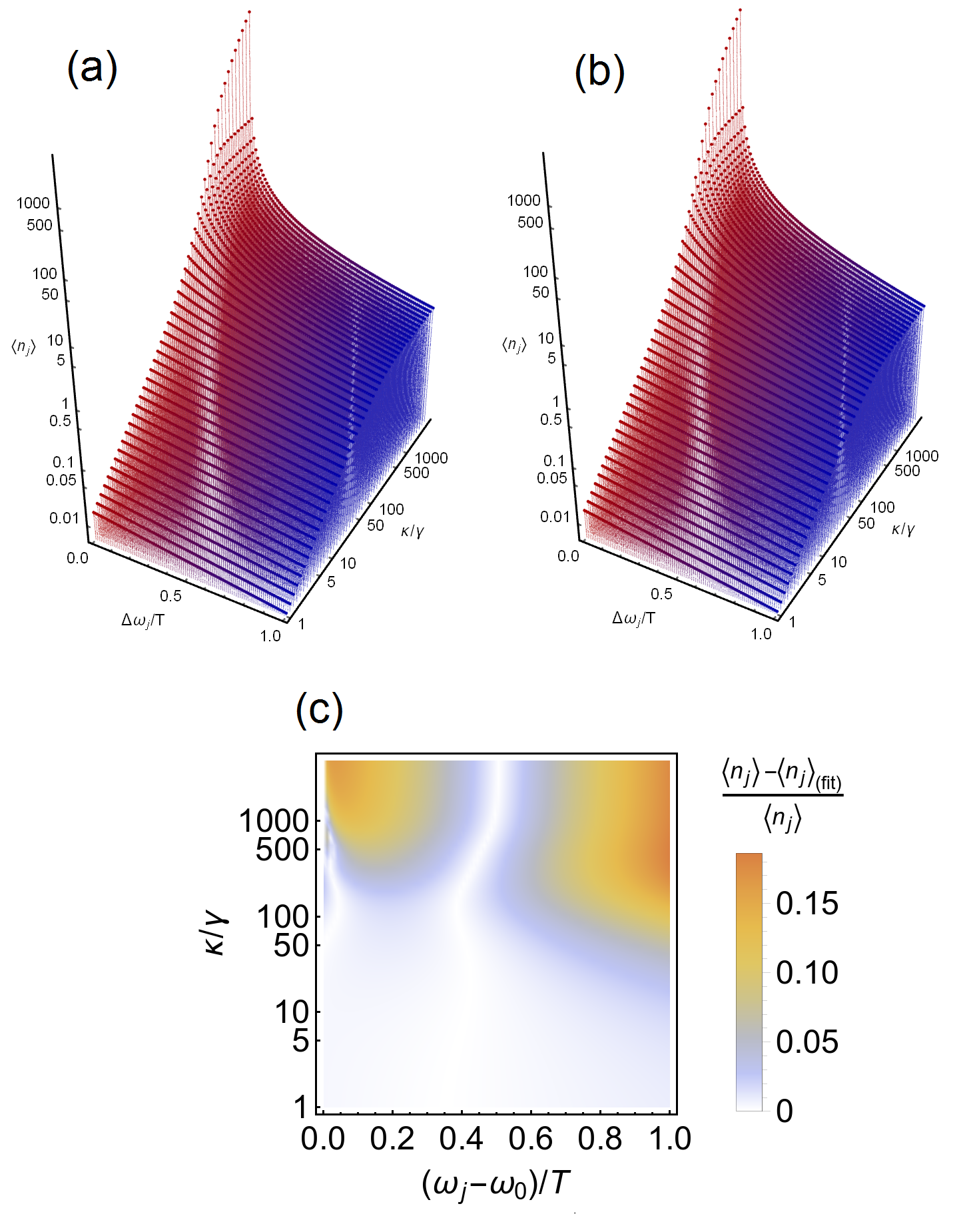}
\caption{\label{fig:BE_approximation} (a) Calculated average population of polaritons. (b) Approximation of the average population of polaritons by the Bose--Einstein distribution with parameter $A = 1.9$ (see~(\ref{BE})). (c) The relative error between the calculated population and the Bose--Einstein distribution.}
\end{figure}

\begin{figure}
\includegraphics[width=0.4\linewidth]{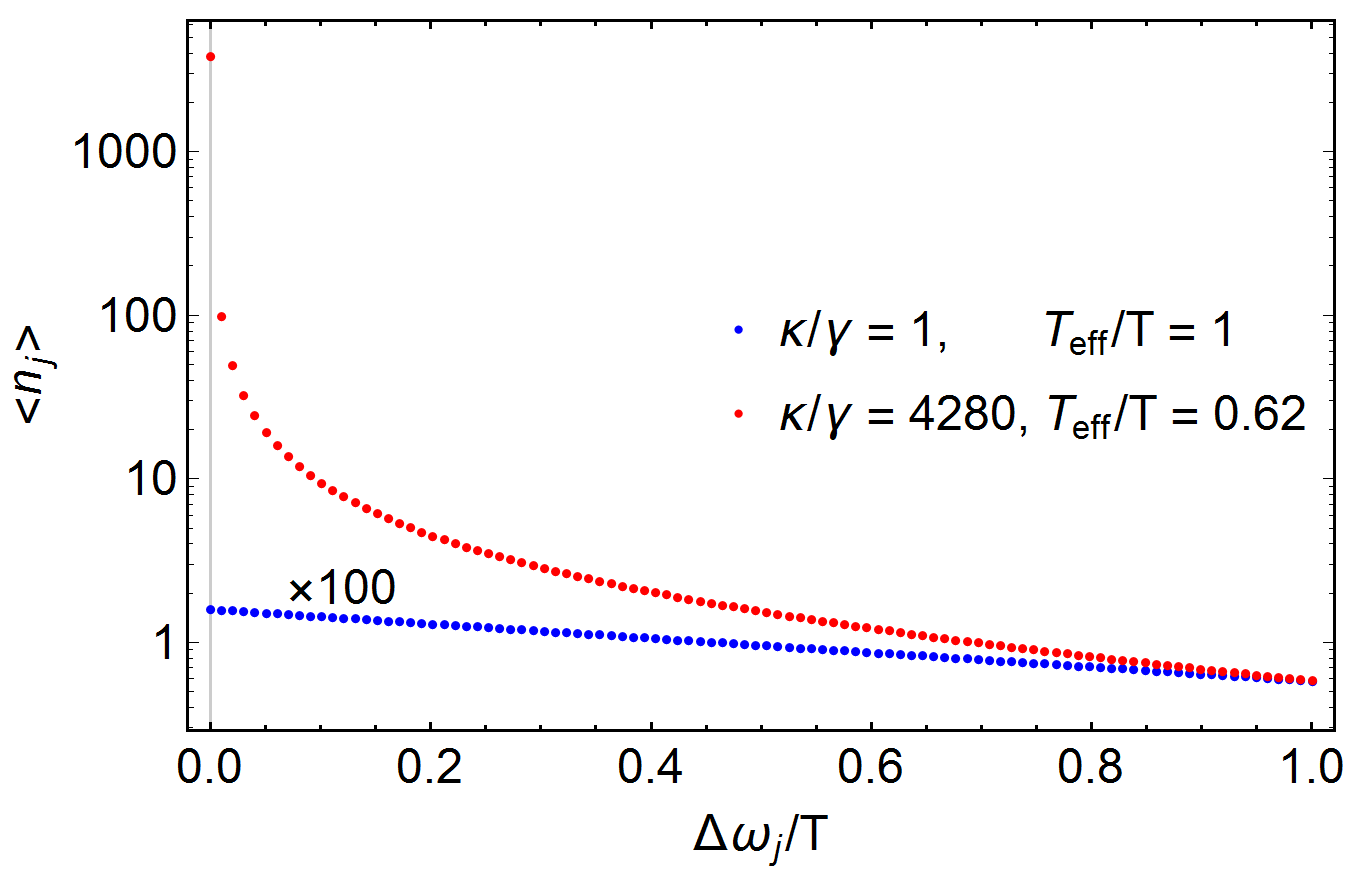}
\caption{\label{fig:Temperature_demonstration} A decrease in the effective temperature of polaritons with incoherent pumping rates above the threshold of the BEC formation.}
\end{figure}

\subsection{\label{sec:SI_6}Dynamics of the formation of non-equilibrium BEC.}
Figure~\ref{fig:Time_dynamics} shows the time dependence of the population of polaritons in the ground state above the threshold.
One can see that the build-up of a stationary population of the ground state occurs in a characteristic time equal to the decay time of polaritons. 
In addition, the higher the rate of the incoherent pumping, the faster the build-up of the stationary population in the ground state.

\begin{figure}
\includegraphics[width=0.4\linewidth]{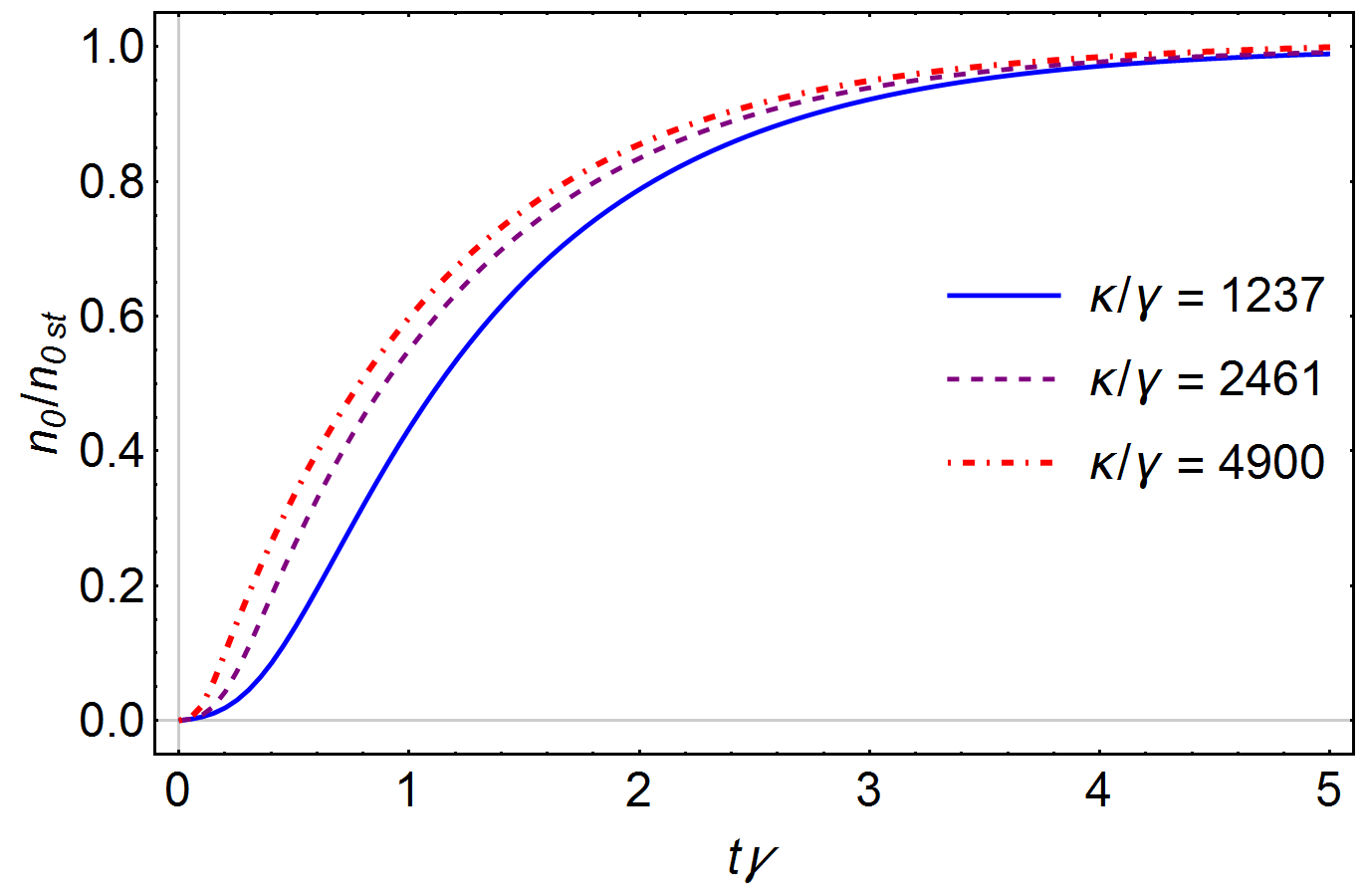}
\caption{\label{fig:Time_dynamics} The ratio of the steady-state population of the ground state and current population of the ground state for pumping rates above the condensate formation threshold.}
\end{figure}

\subsection{\label{sec:SI_7}Formation of polariton BEC when all states of the lower polariton branch are pumped}
Figure~\ref{fig:All_modes} shows the dependence of the average values of polariton population at different pumping rates. 
In this case, incoherent pumping acts on all the polariton states in such a way that its effect on the polaritons can be described by the Lindblad operator~(\ref{Lindblad_pump}) from the main text with ${\kappa _j} = \kappa $ for all polariton states. 
It can be seen from the figure that there are no qualitative differences for the stationary average population of polaritons  between the cases when only the highest-frequency polariton state is pumped and when all polariton states are pumped simultaneously. 

\begin{figure}
\includegraphics[width=0.4\linewidth]{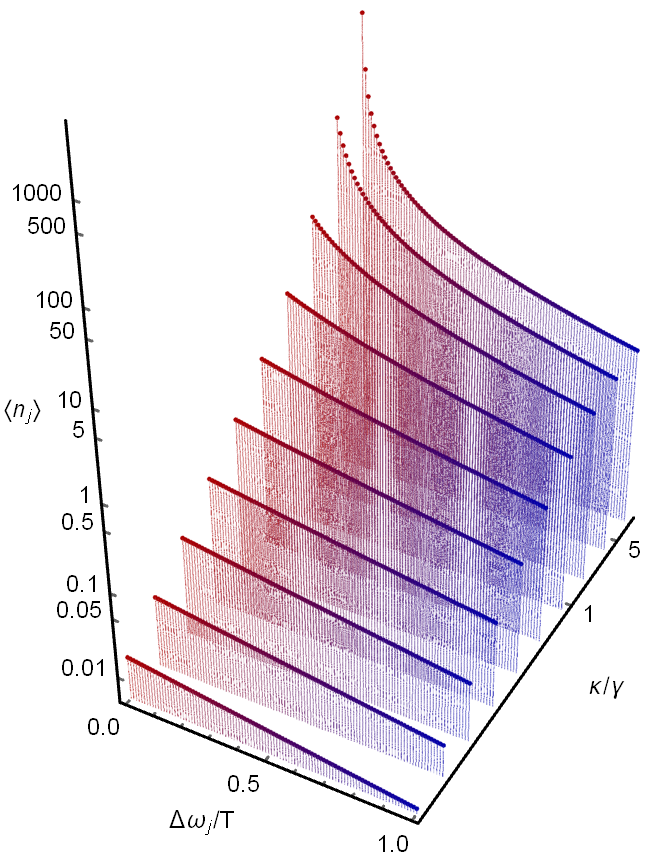}
\caption{\label{fig:All_modes} The average population of lower polaritons when all states are pumped simultaneously.}
\end{figure}
\end{widetext}

\end{document}